\documentclass[10pt, conference, letterpaper]{IEEEtran}

\usepackage{graphicx}
\usepackage{subfigure}
\usepackage[square, comma, sort&compress, numbers]{natbib}
\usepackage{algorithmic}
\usepackage{amsmath}
\usepackage{url}
\usepackage{enumerate}
\usepackage{amssymb}

\begin{document}
	%
	\title{Attention-based Walking Gait and Direction Recognition in Wi-Fi Networks}

	\author{
		\IEEEauthorblockN{Yang Xu$^{\dag}$, Min Chen$^{\ddag}$, Wei Yang$^{\dag}$, Sheng Chen$^{\dag}$, and Liusheng Huang$^{\dag}$\\
			\IEEEauthorblockA{$^{\dag}$School of Computer Science and Technology, University of Science and Technology of China, Hefei, China}
			\IEEEauthorblockA{$^{\ddag}$School of Software Engineering, University of Science and Technology of China, Hefei, China}
		}
	}

	\maketitle
	
	\begin{abstract}
		The study of human gait recognition has been becoming an active research field.
		In this paper, we propose to adopt the attention-based Recurrent Neural Network (RNN) encoder-decoder framework to implement a cycle-independent human gait and walking direction recognition system in Wi-Fi networks.
		For capturing more human walking dynamics, two receivers together with one transmitter are deployed in different spatial layouts.
		In the proposed system, the Channel State Information (CSI) measurements from different receivers are first gathered together and refined to form an integrated walking profile.
		Then, the RNN encoder reads and encodes the walking profile into primary feature vectors.
		Given a specific recognition task, the decoder computes a corresponding attention vector which is a weighted sum of the primary features assigned with different attentions, and is finally used to predict the target.
		The attention scheme motivates our system to learn to adaptively align with different critical clips of CSI data sequence for human walking gait and direction recognitions.
		We implement our system on commodity Wi-Fi devices in indoor environment, and the experimental results demonstrate that our system can achieve average $F_1$ scores of  89.69\% for gait recognition from a group of 8 subjects and 95.06\% for direction recognition from 8 walking directions, in addition, the average accuracies of these two recognition tasks both exceed 97\%.
	\end{abstract}
	

	%
	\IEEEpeerreviewmaketitle

	\section{Introduction}
	Gait can be regarded as an unique feature of a person, and it is usually determined by an individual's physical characters, \textit{e.g.}, height, weight, limb length, and walking habit, \textit{e.g.}, walking speed, posture combined with characteristic motions.
	The study of human gait recognition has been becoming an active research field, especially in the era of Internet of Things (IoT).
	It is appealing that an intelligent house can automatically recognize its owner's gait and offer customized services such as turning on the light and running a bath in advance when the owner comes back outside and walks towards the door, or a special nursing ward can sense an unknown subject in terms of its gait and alert the medical staffs or family members in time.
	Compared with the human identification systems based on other biometrics, like fingerprints, foot pressure, face, iris and voice, which need to be captured by physical contact or at a close distance from the devices, gait-based systems have the potential of unobtrusive and passive sensing \cite{boulgouris2005gait}.
	
	Coincidentally, the emerging Wi-Fi-based human sensing techniques have shown us their potentials of Device-free Passive (DfP) sensing, and have inspired researchers to design and propose many DfP human sensing applications, such as fall detection \cite{han2014wifall}, human daily activity classification \cite{wang2015understanding}, keystroke \cite{ali2015keystroke} and sign language recognition \cite{li2016wifinger,ma2018signfi}.
	The theoretical underpinning of the Wi-Fi-based DfP sensing systems is the Doppler shift and multipath effect of radio signals.
	In Wi-Fi networks, different human activities can induce different Doppler shifts and multipath distortions in Wi-Fi signals, which are depicted and quantized by Channel State Information (CSI).
	For walking movement, the torso and limbs of the walker always move at different speeds, which modulates Wi-Fi signals to the propagation paths with different lengths and introduces different frequency components into the CSI measurements.
	By extracting the very fine-grained and idiosyncratic features from the CSI measurement as human gait representation, some Wi-Fi-based gait recognition systems \cite{wang2016gait,zeng2016wiwho} are proposed.
	Different from the traditional gait recognition systems, which usually rely on video cameras \cite{lee2002gait,lam2011gait}, floor force sensors \cite{orr2000smart}, or wearable devices \cite{sprager2009cumulant,primo2014context}, \textit{etc.} to capture human walking dynamics, Wi-Fi-based systems are unconstrained by light and Line-of-Sight (LoS) conditions, and there is no need to deploy dense specialized sensors or require people to carry or wear some devices.
	Besides, the concern about leaking private data, \textit{e.g.}, image data, is naturally eliminated in the Wi-Fi-based gait recognition systems.
	
	The two critical processes of the existing Wi-Fi based gait recognition systems are gait cycle detection and gait feature extraction \cite{chen2017rapid}.
	However, CSI measurements obtained from commercial Wi-Fi devices contain much noise, which makes it difficult to detect gait cycles.
	Some sophisticated signal processing techniques, like spectrogram enhancement and autocorrelation analysis, are employed to denoise and emphasize the cycle patterns \cite{wang2016gait}.
	After getting the data of each cycle, the previous work proposes to generate some experientially hand-crafted features from time-domain and frequency-domain of the cycle-wise data for gait recognition.
	The previous methods are basically driven by traditional techniques of signal processing and machine learning, which have limitations in extracting high-quality representation of the data. 
	
	In this paper, we propose to adopt a much advanced deep learning framework, namely attention-based Recurrent Neural Network (RNN) encoder-decoder, to implement a cycle-independent human gait and walking direction recognition system with Wi-Fi signals.
	The attention-based RNN encoder-decoder neural networks are initially proposed for machine translation \cite{bahdanau2014neural}. Owing to the attention scheme, the trained networks can adaptively focus the attentions on important words in the source sentence when generating the target word.
	This distinguishing characteristic motivates us to create our cycle-independent gait and direction recognition system given the arbitrarily segmented CSI data.
	Identifying human gait combined with walking direction can enable much more practical and interesting applications while the existing Wi-Fi based gait recognition methods can not cope with these two tasks simultaneously.
	With the attention-based RNN encoder-decoder architecture, the proposed model can jointly train the networks for the two objectives.
	For capturing more human walking dynamics, two receivers and one transmitter are deployed in different spatial layouts.
	In the proposed system, the CSI measurements from the two receivers are first gathered together and refined to form an integrated walking profile.
	Then, the RNN encoder reads and encodes the walking profile into a hidden state sequence, which can be regarded as a primary feature representation of the profile.
	Subsequently, given a specific recognition task (gait or direction), the decoder computes a corresponding attention vector which is a weighted sum of the hidden states assigned with different attentions, and is finally used to predict the recognition target.
	The attention scheme gives the proposed method the ability to align with some critical clips of the input data sequence for the two different tasks.
	The main contributions of this work are summarized as follows:
	\begin{itemize}
		\item By adopting the attention-based RNN encoder-decoder framework, we propose a cycle-independent human gait and walking direction recognition system while the existing Wi-Fi based gait recognition approaches are not reported to cope with these two tasks simultaneously.
		Given a specific recognition task,  the proposed system can adaptively align with the clips, which are critical for that task, of the CSI data sequence.
		To the best of our knowledge, we are among the first to introduce the attention-based RNN encoder-decoder framework to the Wi-Fi based human gait recognition application scenario.
		
		\item In order to capture more human walking dynamics, we deploy two receivers together with one transmitter in different spatial layouts and splice the spectrograms of CSI measurements received by different receivers to construct integrated walking profiles for recognition purpose. 
		A profile reversion method is proposed to augment the training data and our system is trained to cope with multi-direction gait recognition task, thus individuals aren't required to walk along a predefined path in a predefined direction as some existing Wi-Fi-based gait recognition systems do.
		
		\item We implement our system on commodity Wi-Fi devices and evaluate its performance by conducting the walking experiment, where 11 subjects are required to walk on 12 different paths in 8 different directions.
		The experimental results show that our system can achieve average $F_1$ scores of 89.69\% for gait recognition from a group of 8 randomly selected subjects and 95.06\% for direction recognition from 8 directions, and the average accuracies of these two recognition tasks both exceed 97\%.
	\end{itemize}
	
	\section{Related Work}
	In the 1970's, Johansson et al. \cite{johansson1973visual} and Cutting et al. \cite{cutting1978biomechanical} had conducted similar research and found that viewers could determine the gender of a walker or even recognize the walker who they were familiar by just watching the video pictures of prominent joints of the walker.
	Based on these study foundations, the early human walking activity and gait recognition applications were mainly based on video or image sequences \cite{polana1994detecting,little1998recognizing,nixon1999automatic,lee2002gait,boulgouris2005gait,lam2011gait}.
	Polana et al. \cite{polana1994detecting} proposed to recognize human walking activity by analyzing the periodicity of different activities in optical flow.
	Besides periodicity, Little et al. \cite{little1998recognizing} also extracted moment features of foreground silhouettes and optical flow for walking identification.
	Moreover, Lee et al. \cite{lee2002gait} divided the silhouettes into 7 regions that roughly correspond to different body parts and computed statistics on these regions to construct gait representation and realize human identification and gender classification.
	And some other useful methods like Gait Energy Image (GEI) \cite{han2006individual} and Gait Flow Image (GFI) \cite{lam2011gait} were proposed to further enhance the equality of gait representation created from silhouettes and improve the rate of gait recognition.
	However, the video-based methods always require subjects to walk in the direction perpendicular to the optical axis of cameras so as to get more gait information \cite{boulgouris2005gait}, and also introduce many other non-negligible drawbacks, such as LoS condition, light condition and personal privacy.
	Besides, some other data sensors were employed for gait recognition.
	Orr et al. \cite{orr2000smart} adopted floor force sensors to record individuals' footstep force profiles, based on which the footstep models were built and achieved an overall gait recognition rate of 93\% for identifying a single subject.
	Sprager et al. \cite{sprager2009cumulant} and Primo et al. \cite{primo2014context} proposed to use the built-in accelerometers of mobile phones to collect walking dynamics and extract gait features for human recognition.
	These sensor-based approaches either rely on specially deployed sensors (\textit{e.g.}, floor sensors) or require people to carry or wear wearable devices, which limits their applicability.
	
	Recently, the emerging Wi-Fi-based (mainly, CSI-based) sensing techniques are widely applied to produce many DfP applications, such as human activity recognition \cite{han2014wifall, wang2015understanding, ali2015keystroke, li2016wifinger,ma2018signfi}, indoor localization \cite{sen2012you}, gait recognition \cite{zeng2016wiwho, wang2016gait}.
	Since we mainly concern the problem of using Wi-Fi CSI to implement gait recognition, here we introduce some previous work on CSI-based gait recognition.
	By utilizing the time-domain and frequency-domain gait features extracted from CSI, Zeng et al. \cite{zeng2016wiwho}, Zhang et al. \cite{zhang2016wifi} and Wang et al. \cite{wang2016gait} separately proposed three gait recognition systems called WiWho, WiFi-ID and WifiU.
	WiWho mainly focused on the low frequency band 0.3$\sim$2 Hz of CSI, which contains much interference induced by slight body movements and environment changes \cite{xu2018wistep}.
	This hinders WiWho from working when the subject is as far as more than 1 meter from the LoS path of its sender and receiver.
	While WiFi-ID and WifiU concentrated on the frequency components of 20$\sim$80 Hz in CSI measurements.
	In WiFi-ID, Continuous Wavelet Transformation (CWT) and RelieF feature selection algorithm were applied to extract gait features in different frequency bands, and the Sparse Approximation based Classification (SAC) \cite{wright2009robust} was chosen as the classifier.
	In WifiU, Principal Component Analysis (PCA) and spectrogram enhancement techniques were utilized to generate a synthetic CSI spectrogram from which a set of 170 features were derived, and the SVM classifier was finally employed for human identification.
	Based on WiWho, Chen et al. \cite{chen2017rapid} introduced an acoustic sensor (\textit{i.e.}, a condenser microphone) as a complementary sensing module to implement a multimodal human identification system, named Rapid, which could guarantee a more robust classification result in comparison to WiWho.
	Most of these systems could achieve an average human identification accuracy of around 92\% to 80\% from a group of 2 to 6 subjects.
	And one important function of the microphone used in Rapid is to detect the start and end points of each walking step, \textit{i.e.}, gait cycle detection, which is an indispensable part of the vast majority of video-, sensor- and CSI-based gait recognition systems.
	
	However, segmenting gait cycles from CSI measurements is difficult since the variation patterns induced by walking are sometimes buried in the noise \cite{wang2016gait}, and some sophisticated signal processing techniques are needed to carefully refine the data.
	After gait cycle detection, most existing methods (like Rapid, WifiU, \textit{etc.}) try to generate some experientially hand-crafted features from the cycle-wise gait data and then train the gait classifiers.
	Different from all the aforementioned work, our system introduces the attention-based RNN encoder-decoder architecture to (i) adaptively align with some important time slices of CSI data sequence, which means there is no cycle partitioning needed, (ii) automatically learn to extract effective feature representations for gait recognition.
	Another major difference between our system and the above work is that our system is trained to cope with multi-direction gait recognition (12 walking paths and 8 walking directions relative to the system equipments), where individuals don't have to walk along a predefined path in a predefined direction as WiWho or WifiU does.
	In what follows, we will specify the design of the proposed system.
	
	\section{Background \& Motivation}
	To understand how human walking activity exerts impacts on Wi-Fi signals, we first give a brief overview of CSI and the multipath effect of wireless signals.
	And then, we explain the network architecture of the attention-based RNN encoder-decoder, which powers our system to run the core functions.
	
	\subsection{Channel State Information}
	As for wireless communication, Channel Frequency Response (CFR) characterizes the small-scale multipath effect and the frequency-selective fading of the wireless channels.
	Let $X(f,t)$ and $Y(f,t)$ separately denote the frequency-domain transmitted and received signal vectors, and the relation between $X$ and $Y$ can be modeled as
	$Y(f, t) = H(f, t)\times X(f, t) + \mathcal{N}_{\emph{noise}}$,
	where $f$ is the carrier frequency, $t$ indicates that the channel is time-varying, $H(f,t)$ is the complex valued CFR and $\mathcal{N}_{\emph{noise}}$ represents the noise.
	Furthermore, in Wi-Fi networks, Channel State Information (CSI) is used to monitor the channel properties, and it is the simple version of CFR with discrete subcarrier frequency $f_i$ ($i=1, 2,..., N_S$), where $N_S$ denotes the total number of subcarriers across an Wi-Fi channel.
	With the CSI tool \cite{halperin2011tool}, an CSI measurement, which contains 30 matrices with dimensions $N_{\emph{Tx}}\times N_{\emph{Rx}}$, can be got from one physical frame, $N_{\emph{Tx}}$ and $N_{\emph{Rx}}$ represent the number of antennas of the transmitter (Tx) and the receiver (Rx), respectively.
	We regard the CSI measurements collected from each Tx-Rx antenna pair as a \emph{CSI stream}, thus there are $N_C = N_{\emph{Tx}}\times N_{\emph{Rx}}$ time-series CSI streams.
	
	\begin{figure}[!t]
		\centering
		\includegraphics[width=0.75\columnwidth]{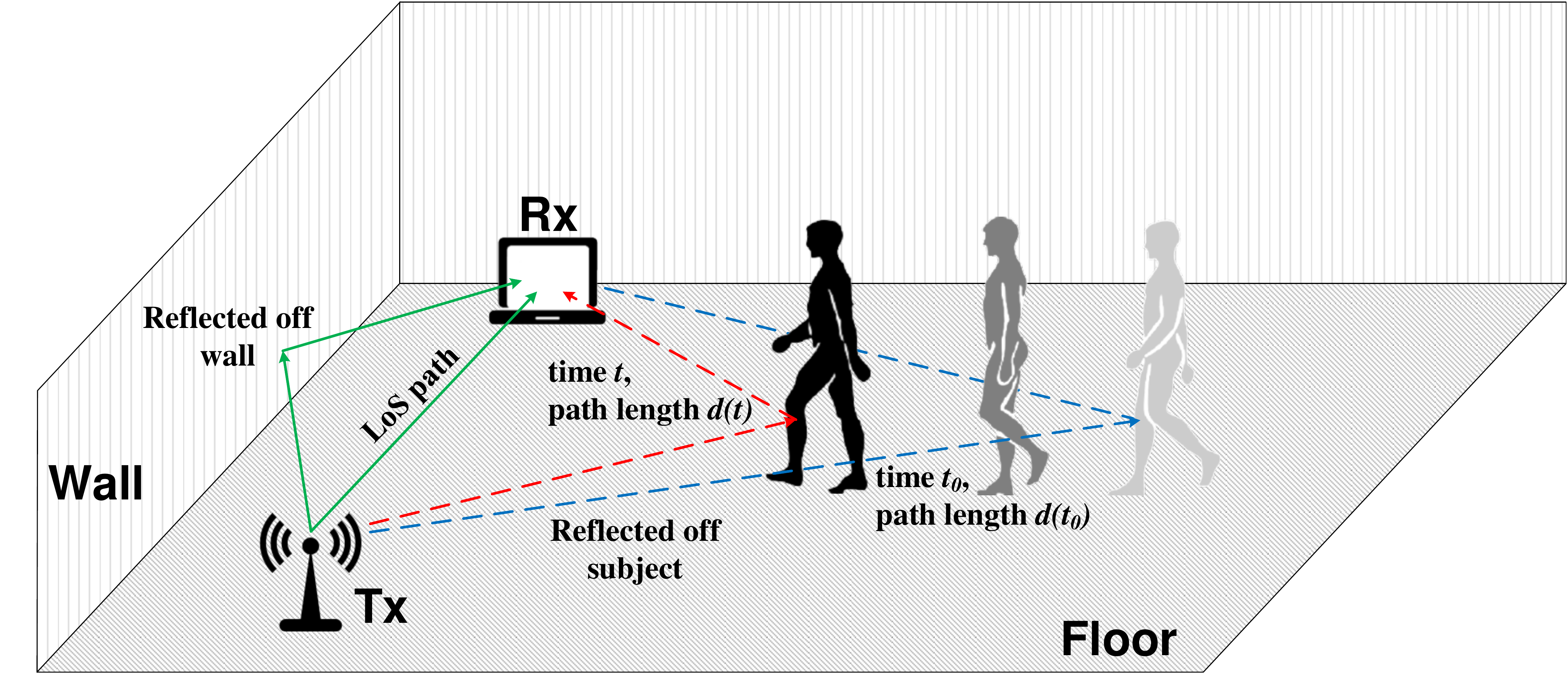}
		\caption{Multipath effect of Wi-Fi signals in indoor environment.}
		\label{fig_multipath_effect}
	\end{figure}
	
	\subsection{Multipath Effect of Wi-Fi Signals}
	In Fig. \ref{fig_multipath_effect},	one transmitted signal can directly travel through the Line-of-Sight (LoS) path, or be reflected off the wall and the walking subject and propagate through multiple paths before arriving at the receiver, this phenomenon is called \emph{multipath effect} \cite{rappaport1996wireless}.
	The multiple paths can be divided into two categories: static paths (solid lines in Fig. \ref{fig_multipath_effect}) and dynamic paths (caused by walking subject as expressed by dotted lines in Fig. \ref{fig_multipath_effect}).
	If there totally exist $N_P$ propagation paths among which $N_D$ paths are dynamically varying, then $H(f_i, t)$ of the $i^{th}$ subcarrier at time $t$ can roughly be expressed as
	\begin{equation}\label{eqn_multipath}
	\begin{aligned}[b]
	H(f_i,t) =& \sum_{n=1}^{N_P} a_n(f_i,t)e^{-j\frac{2\pi d_n(t) }{\lambda_i}}\\
	=& H_s(f_i) + \sum_{k=1}^{N_D} a_k(f_i,t)e^{-j\frac{2\pi d_k(t) }{\lambda_i}}\mbox{,}
	\end{aligned}
	\end{equation}
	where $H_s(f_i)$ is the sum of responses of the static paths and can be regarded as a constant \cite{wang2015understanding}, since signals traveling through the static paths have relatively invariable path length and propagation attenuation,
	$a_k(f_i,t)$ represents the attenuation of the $k^{th}$ dynamical path, $\frac{d_k(t) }{\lambda_i}$ and $\frac{2\pi d_k(t) }{\lambda_i}$ separately denote the propagation delay and the phase shift when the path length is $d_k(t)$, $\lambda_i$ is the wavelength of subcarrier $i$.
	In terms of Fig. \ref{fig_multipath_effect}, at time $t$, the length of path $k$ can be expressed as $d_k(t) = d_k(t_0) + v_kt$, where $v_k$ represent change speed of path $k$.
	Therefore, $\frac{2\pi d_k(t) }{\lambda_i} = \frac{2\pi v_kt }{\lambda_i} + \frac{2\pi d_k(t_0) }{\lambda_i}$.
	Usually, a $\lambda_i$ displacement of the subject can roughly cause a $2\lambda_i$ length change of the dynamical path $k$ (round trip), which introduces $4\pi$ phase change.
	According to the principle of superposition of waves, the $4\pi$ phase change finally induces 2 cycles of the amplitude change of CSI values \cite{wang2015understanding}, which reveals an approximate relation between human moving speed $V$ ($= v_k/2$) and frequency $F$ of amplitude variation of CSI values, \emph{i.e.}, $F = 2V / \lambda_k$.
	
\begin{figure}[!t]
		\centering
		\subfigure[The variation of CSI amplitude when a person walks away from the Tx and the Rx.]
		{
			\includegraphics[width=0.45\columnwidth]{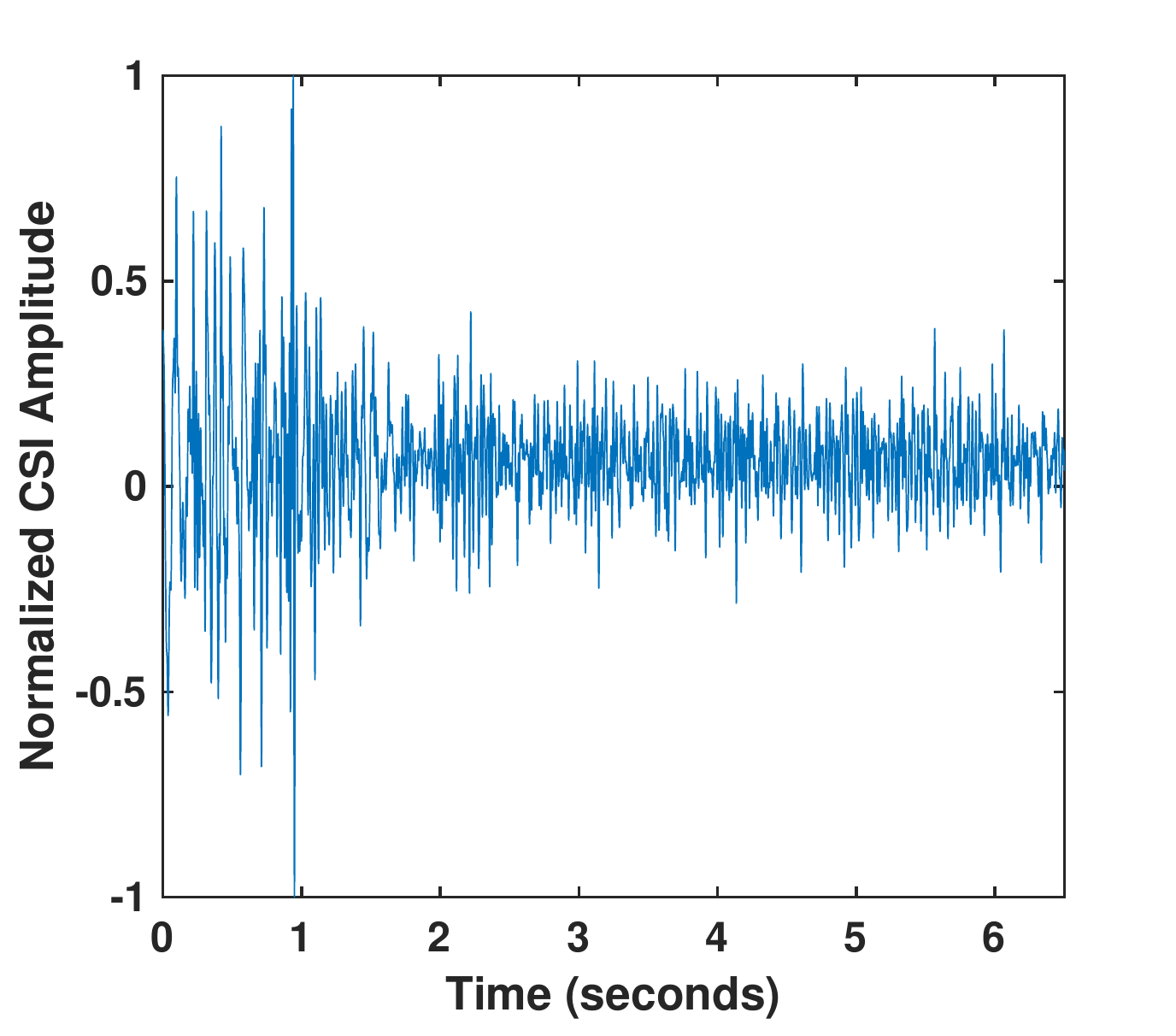}
			\label{fig_walk_away_waveform}
		}\quad
		\subfigure[The variation of CSI amplitude when a person walks towards the Tx and the Rx.]
		{
			\includegraphics[width=0.45\columnwidth]{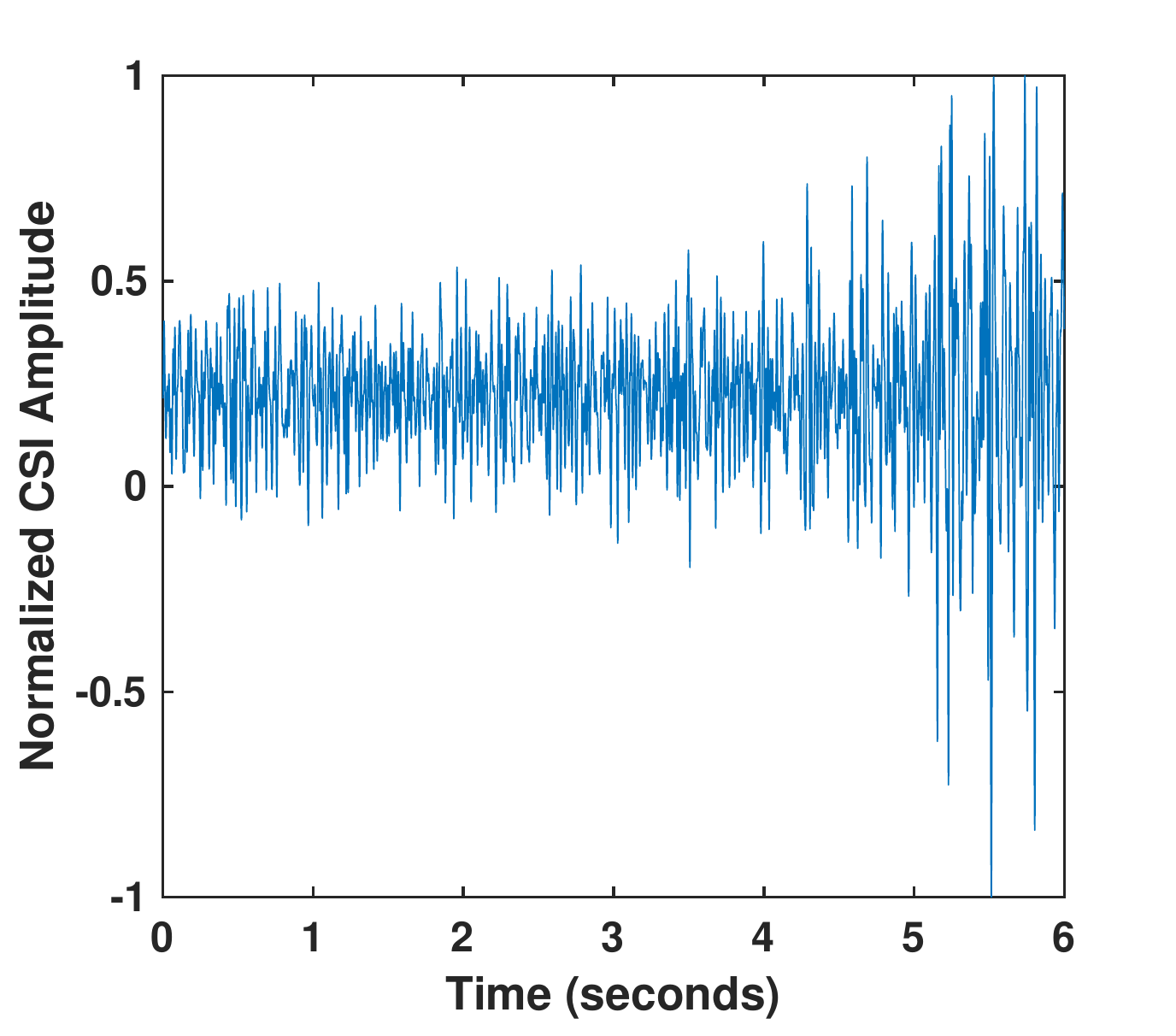}
			\label{fig_walk_close_waveform}
		}
		\caption{The variation of CSI amplitude when a person walks in different directions relative to the Tx and the Rx (in Fig. \ref{fig_multipath_effect}).}
		\label{fig_walking_instance_waveform}
	\end{figure}

	In addition, according to the Friis free space propagation model, the transmitting power ($P_t$) and the receiving power ($P_r$) of Wi-Fi signals have the following relation \cite{rappaport1996wireless}:
	\begin{equation}\label{eqn_friis}
	P_r = P_t G_r G_t \big(\frac{\lambda}{4 \pi d}\big)^2 \mbox{,}
	\end{equation}
	$G_r$ and $G_t$ separately are the gains of Rx and Tx antennas, $d$ is the length of LoS path.
	Besides, the receiving power ($P_r^i$) of the $i^{th}$ subcarrier is proven to be basically proportional to the CSI power ($|H(f_i,t)|^2$) of subcarrier $i$ \cite{yang2013rssi,wu2012fila}, \textit{i.e.}, $P_r^i \propto |H(f_i,t)|^2$.
	Thus, combined with equation (\ref{eqn_friis}), we can get the relation $|H(f_i,t)|^2 \propto \frac{1}{d^2}$, 
	
	Based on the above explanation, we can get some useful information:
	\begin{enumerate}
		\item Since every individual has unique gait while walking, which means different people can induce Wi-Fi signals propagating through different paths and result in quite different changing patterns of the paths.
		Fortunately, these different patterns can be imprinted on CSI measurements and be reflected by the changing patterns of CSI amplitudes and we can probably realize gait recognition task by digging into the CSI measurements.
		\item As shown in Fig. \ref{fig_multipath_effect}, assuming that Tx and Rx are fixed, \textit{i.e.}, the distance of the LoS path is constant.
		If there is no moving object in the environment, the power of the received CSI measurements will be relatively steady.
		However, when a person walks towards the Tx and the Rx, the distances of static paths are still constant while the distances of dynamic paths induced by the person is getting shorter, and the receiving power of the dynamic paths is getting higher, thus the variance of CSI power or CSI amplitude becomes larger and larger, and vice versa.
		Fig. \ref{fig_walking_instance_waveform} displays the variation of CSI amplitude when a person walks in different directions relative to the Tx and the Rx.
		Therefore, we can roughly regard that $\sigma^2(|H(f_i,t)|^2) \propto \frac{1}{d^2}$, and \cite{chen2017rapid} also drew the similar conclusion.
		Based on this, we can deduce the walking direction of a person by analyzing the variation trend of CSI power or CSI amplitude. 
	\end{enumerate}
	
	By now, we have explained the basis and feasibility of walking gait and direction recognition using Wi-Fi signals.
	Next, we will introduce the core model employed in the proposed method, namely the attention-based RNN encoder-decoder.
	
	\subsection{Attention-based RNN encoder-decoder}
	
	\begin{figure}[!t]
		\centering
		\includegraphics[width=0.90\columnwidth]{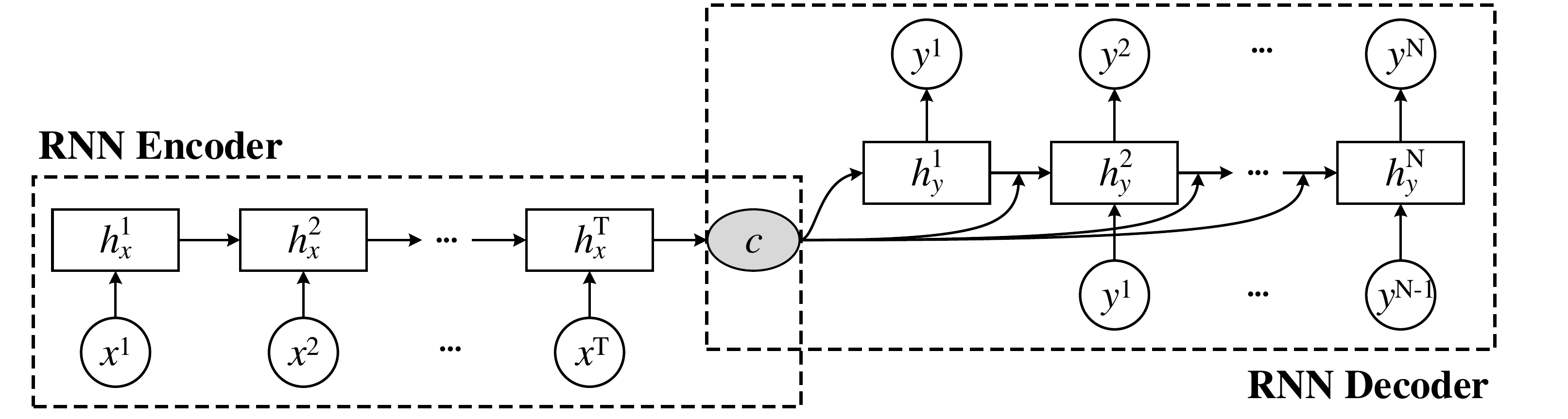}
		\caption{The standard architecture of RNN encoder-decoder.}
		\label{fig_standard_RNN_encoder_decoder}
	\end{figure}
	
	\subsubsection{Standard RNN encoder-decoder}
	The time-series CSI measurements are a kind of sequence data whose lengths can be arbitrary, given a sequence of CSI inputs ($x^1, \cdots, x^T$), the target of our system is to output another predicted sequence ($y^1, \cdots, y^N$), such as walking direction and walker identity.
	Therefore, this process can be formulated as a sequence to sequence learning problem \cite{sutskever2014sequence}, and the RNN encoder-decoder is naturally adopted in this work.
	Fig. \ref{fig_standard_RNN_encoder_decoder} illustrates the architecture of a standard RNN encoder-decoder \cite{cho2014learning}, which learns to encode a input sequence into a fixed-length summary vector $c$ and to decode the vector $c$ into an output sequence.
	At each time step $t$, the hidden state $h_x^t$ of the RNN encoder is updated by
	\begin{equation}\label{eqn_encoder_hidden_state}
	h_x^t = f(h_x^{t-1}, x^t) \mbox{,}
	\end{equation}
	and the summary vector $c$ is generated by
	\begin{equation}\label{eqn_summary_vector}
	c = q(h_x^1, \cdots, h_x^T) \mbox{,}
	\end{equation}
	where $f$ is a non-linear activation function, and it can be a Long Short-Term Memory (LSTM) unit \cite{hochreiter1997long} or a Gated Recurrent Unit (GRU) \cite{cho2014learning}, which can automatically extract high-quality features of the input data, and $q$ can be a simple function of picking the last hidden state, \textit{i.e.}, $q(h_x^1, \cdots, h_x^T) = h_x^T$.
	
	The decoder is another RNN, and it is trained to learn the following conditional distribution:
	\begin{equation}\label{eqn_output_conditional_distribution}
	P(y^n|y^{n-1}, y^{n-2}, \cdots, y^1, c) = g(h_y^n, y^{n-1}, c), n \in [1, N] \mbox{,}
	\end{equation}
	where
	\begin{equation}\label{eqn_decoder_hidden_state}
	h_y^n = f(h_y^{n-1}, y^{n-1}, c) \mbox{,}
	\end{equation}
	here, $y^n$ is the $n^{th}$ predicted result, $h_y^n$ is the hidden state of the $n^{th}$ prediction step, $f$ also is a non-linear function and $g$ usually is a softmax function.
	\textit{With the help of RNN encoder-decoder model, the proposed system can jointly train to maximize the probability of a series of recognition tasks (\textit{e.g.}, direction estimation and human identification) given an CSI sequence}.
	
	However, a major concern is that the standard RNN encoder-decoder model tries to compress all input data into a single fixed-length vector, which is used to predict all the output labels.
	Since different outputs probably have different connections to the inputs, for example, prediction of walking direction may need to sense the variation trend of the entire CSI power sequence while  human identification may focus more on some critical clips of the CSI sequence.
	To address this concern, the attention-based RNN encoder-decoder is adopted in our method.
	
	\begin{figure}[!t]
		\centering
		\includegraphics[width=0.9\columnwidth]{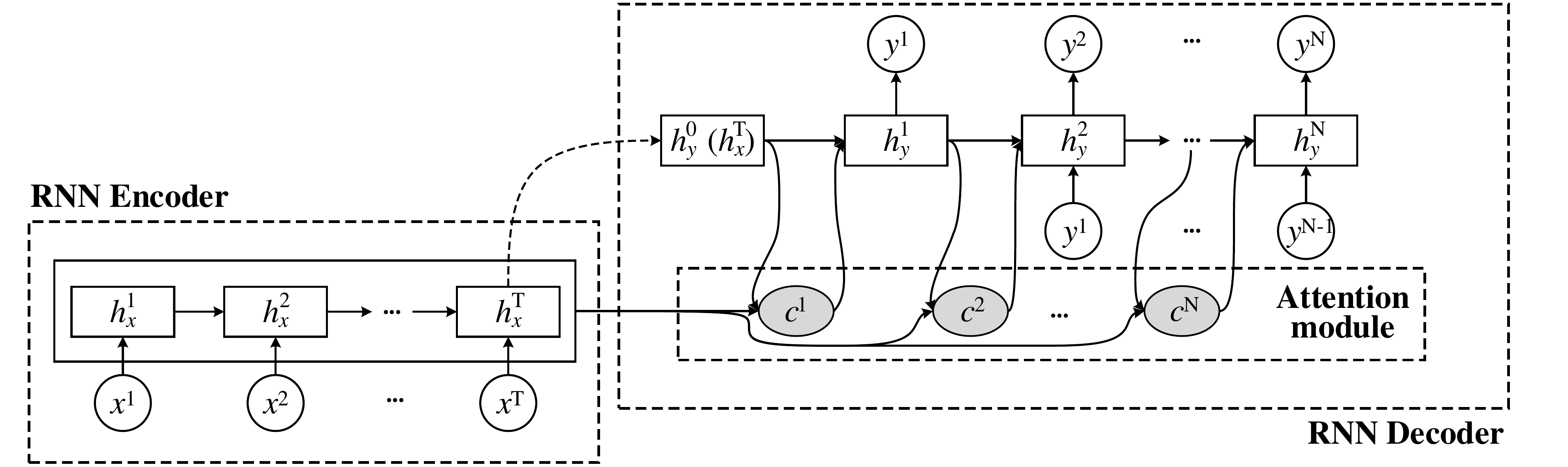}
		\caption{The architecture of attention-based RNN encoder-decoder.}
		\label{fig_attention_based_RNN_encoder_decoder}
	\end{figure}
	
	\subsubsection{RNN encoder-decoder with attention scheme} \label{subsubsec_attn_rnn_en_de}
	As Fig. \ref{fig_attention_based_RNN_encoder_decoder} shows, the key difference between attention-based and standard RNN encoder-decoder is that the attention-based model adaptively encodes the input sequence into different summary vectors, which we call attention vectors, for different predictions.
	The attention-based RNN encoder-decoder model is firstly proposed for machine translation \cite{bahdanau2014neural}, and it can learn to align and translate simultaneously.
	In this model, the conditional probability of equation (\ref{eqn_output_conditional_distribution}) is rewritten as
	\begin{equation}\label{eqn_attention_output_conditional_distribution}
	P(y^n|y^{n-1}, y^{n-2}, \cdots, y^1, c^n) = g(h_y^n, y^{n-1}, c^n), n \in [1, N] \mbox{,}
	\end{equation}
	and $h_y^n = f(h_y^{n-1}, y^{n-1}, c^n)$.
	The derivation of distinct attention $c_n$ corresponding to the target $y^n$ is expressed as the weighted sum of all the hidden states of RNN encoder ($h_x^1, \cdots, h_x^T$):
	\begin{equation}\label{eqn_summary_vector_c_n}
	c^n = \sum_{t = 1}^{T}\omega^{n,t} h_x^t \mbox{,}
	\end{equation}
	where $\omega^{n,t}$ is the attention weight, and it is computed by
	\begin{equation}\label{eqn_attention_weight}
	\omega^{n,t} = softmax(attn(h_y^n,h_x^t)) \mbox{,}
	\end{equation}
	the function $attn(\cdot)$ scores how well the input data at each time step and the current output match \cite{bahdanau2014neural}, which enables the model to adaptively align with some important parts of the inputs when predicting a certain target.
	It's is noticed that the initial hidden state $h_y^0$ of the RNN decoder is set as $h_x^T$ in the specific implementation of \cite{bahdanau2014neural}.

	\section{System Design}\label{sec_system_design}
	The proposed system consists of four essential modules, which are \textit{CSI Collection}, \textit{Raw CSI Processing}, \textit{Walking Profile Generation} and \textit{Gait and Direction Recognition}.
	In what follows, the detailed processing procedures of each module will be explained.
	
	\subsection{CSI Collection} \label{subsec_csi_collection}
\begin{figure}
		\centering
		\subfigure[Amplitude variance of in-place walking w/o swing arms.]
		{
			\includegraphics[width=0.75\columnwidth]{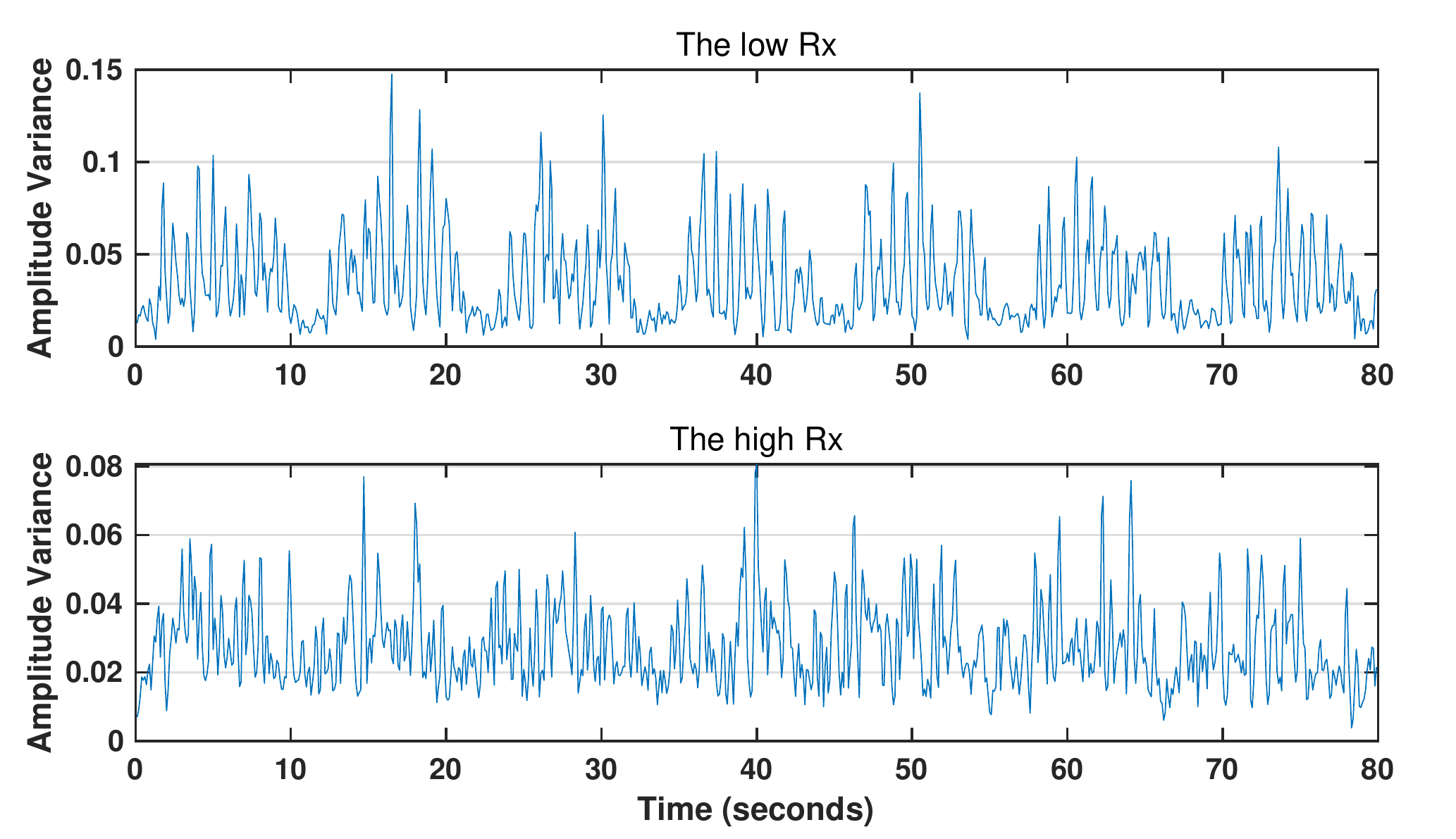}
			\label{fig_var_walking_wo_arm}
		}\\
		\subfigure[Amplitude variance of swing arms w/o walking.]
		{
			\includegraphics[width=0.75\columnwidth]{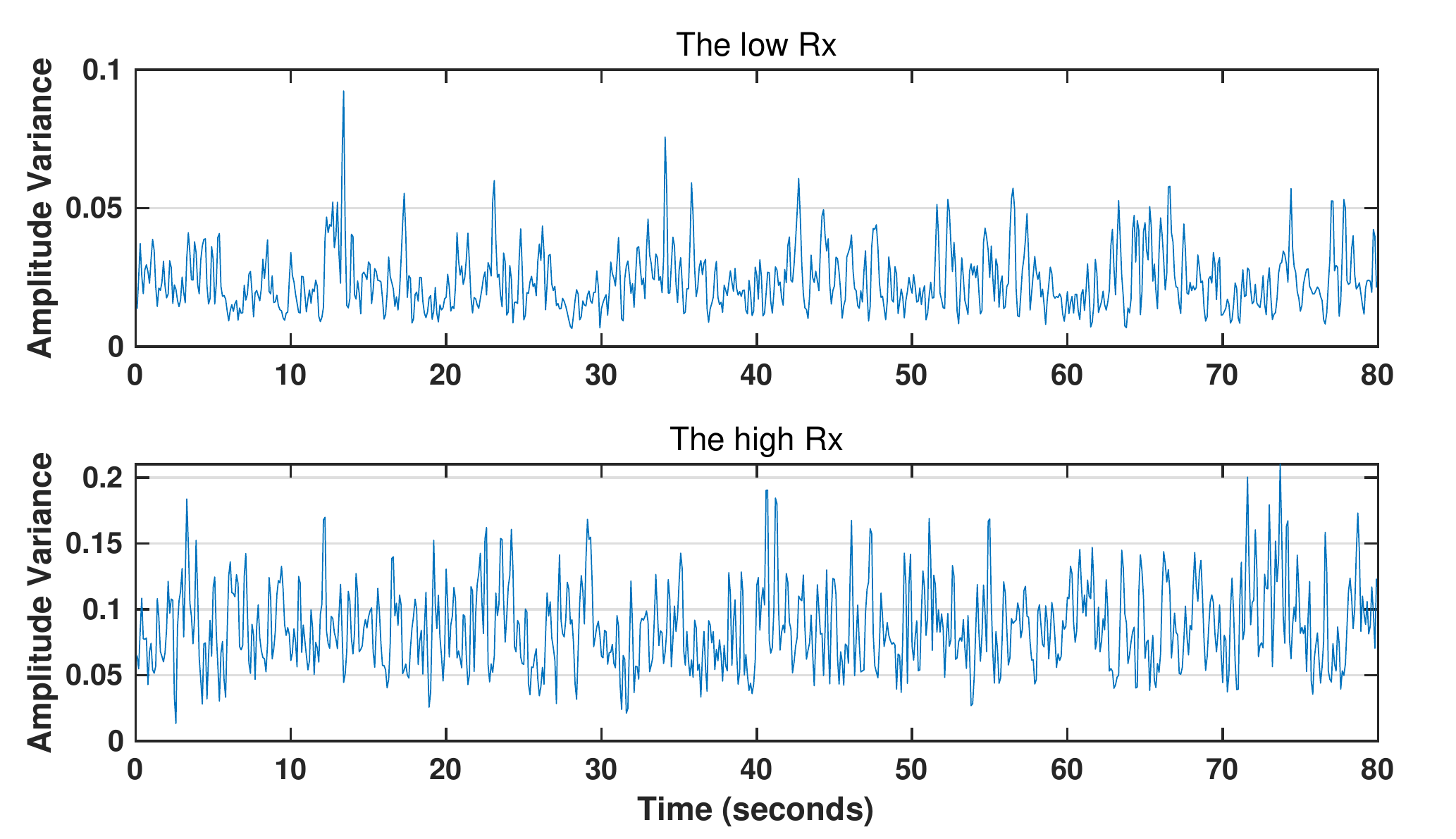}
			\label{fig_var_swing_arm}
		}
		\caption{The amplitude variance of CSI received by different receivers.}
		\label{fig_csi_amplitude_variance}
	\end{figure}

	Since the 2.4 GHz Wi-Fi band is narrower and more crowded than the 5 GHz band, the latter is a much better choice for less inter-channel interference and more reliable communication.
	Therefore, our system is set to run on the 5 GHz Wi-Fi band.
	A laptop equipped with an Intel 5300 wireless card and 2 omni-directional antennas serves as the transmitter.
	In order to capture more walking dynamics, two laptops equipped with Intel 5300 wireless cards (each with 3 omni-directional antennas) are deployed as the receivers.
	For concentrating on different body parts of a walker, the receivers are placed at different heights, where one is at a height of 0.5 m and the other is 1.0 m above the ground level, and the transmitter is placed at a height of 0.75 m.
	The transmitter continuously sends 802.11n data packets to the receivers, to which the CSI measurements of correctly received packets are reported with the tool released in \cite{halperin2011tool}.
	Fig. \ref{fig_csi_amplitude_variance} illustrates the amplitude variance of CSI received by the 2 receivers when a subject performs two different movements (in-place walking w/o swing arms and swing arms w/o walking), and we can find that the lower receiver is more sensitive to leg movements while the higher is more sensitive to arm movements.
	Considering human activities in traditional indoor environment introduce frequencies of no more than 300 Hz in CSI measurements \cite{wang2015understanding}, in terms of the Nyquist sampling theorem,  our system is configured with a sampling rate ($F_s$) of 1000 Hz.
	For each receiver, the received data of each CSI stream forms a matrix $\textbf{C}_i (i=1 \cdots N_C)$ with dimensions $N_S \times T$, where $N_C=6\ (2 \times 3)$ and $N_S=30$ separately denote the number of streams and subcarriers in each stream.
	$T$ is the data length.
	To eliminate the impact of Carrier Frequency Offset (CFO), we only reserve the CSI amplitude and ignore the CSI phase in our system as \cite{wang2015understanding} suggested.

	\subsection{Raw CSI Processing}
	
	\subsubsection{Long Delay Removal}
	Channel Impulse Response (CIR), which is the inverse Fourier transformation of CFR, can characterize the propagation delays of the received signals.
	The signals with long propagation delay probably are reflected by some static or dynamic objects which are far away from the transceivers, and these signals are useless and can distort the CSI amplitudes.
	Theoretically, every signal with a certain propagation delay can be separated from CIR, but limited by the bandwidth of Wi-Fi channel (\textit{i.e.}, 20 MHz), the time resolution of CIR is approximately 1$/$20MHz $=$ 50 ns \cite{xie2015precise}.
	Therefore, we can only distinguish a series of signal clusters with discrete time delays.
	Besides, previous study shows the maximum delay in general indoor environment is less than 500 ns \cite{jin2010indoor}.
	Thus, we transform each CSI measurement into time-domain CIR by Inverse Fast Fourier transformation (IFFT) and remove the components whose propagation delays are longer than 500 ns, and then we convert the processed CIR back to CSI by Fast Fourier Transformation (FFT).
	
	\subsubsection{CSI Denoising}
	After removing the paths with long delays, the CSI values still contain significant high-frequency noise and low-frequency interferences \cite{xu2018wistep}.
	Moreover, the frequency components, \textit{i.e.}, the frequencies of CSI amplitude variation, induced by walking are approximately within the range of 20$\sim$60 Hz given the 5 GHz Wi-Fi band\cite{wang2016gait,xu2018wistep},
	the proposed system adopts the Butterworth bandpass filter, which guarantees high fidelity of reserved signals in the passband, to eliminate the high-frequency and low-frequency noise.
	The upper and lower cutoff frequencies of the Butterworth filter are empirically set as 90 Hz and 5 Hz, respectively.
	The direct current component (0 Hz) of each subcarrier is also filtered by the bandpass filter.
	Subsequently, the proposed system introduces weighted moving average to further denoise and smooth the CSI amplitudes.
	
	\subsubsection{CSI Refining}
\begin{figure}[!t]
		\centering
		\subfigure[]
		{
			\includegraphics[width=0.45\columnwidth]{./figures/subject1_dd+_pwr_d0_seg_2.pdf}
			\label{fig_pwr_d0_waveform}
		}
		\subfigure[]
		{
			\includegraphics[width=0.45\columnwidth]{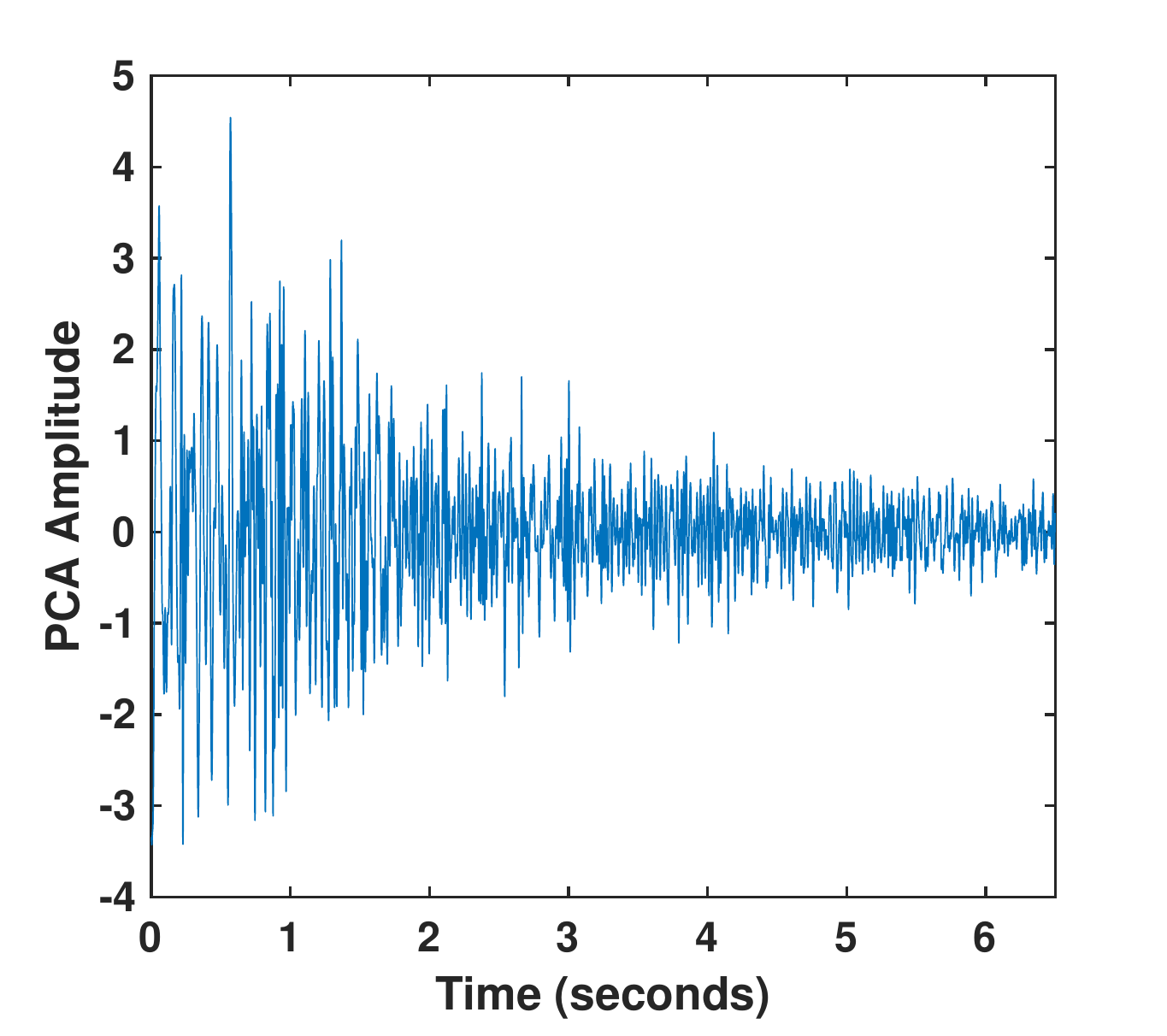}
			\label{fig_pca_d0_waveform}
		}\\
		\subfigure[]
		{
			\includegraphics[width=0.45\columnwidth]{./figures/subject1_dd+_pwr_d1_seg_3.pdf}
			\label{fig_pwr_d1_waveform}
		}
		\subfigure[]
		{
			\includegraphics[width=0.45\columnwidth]{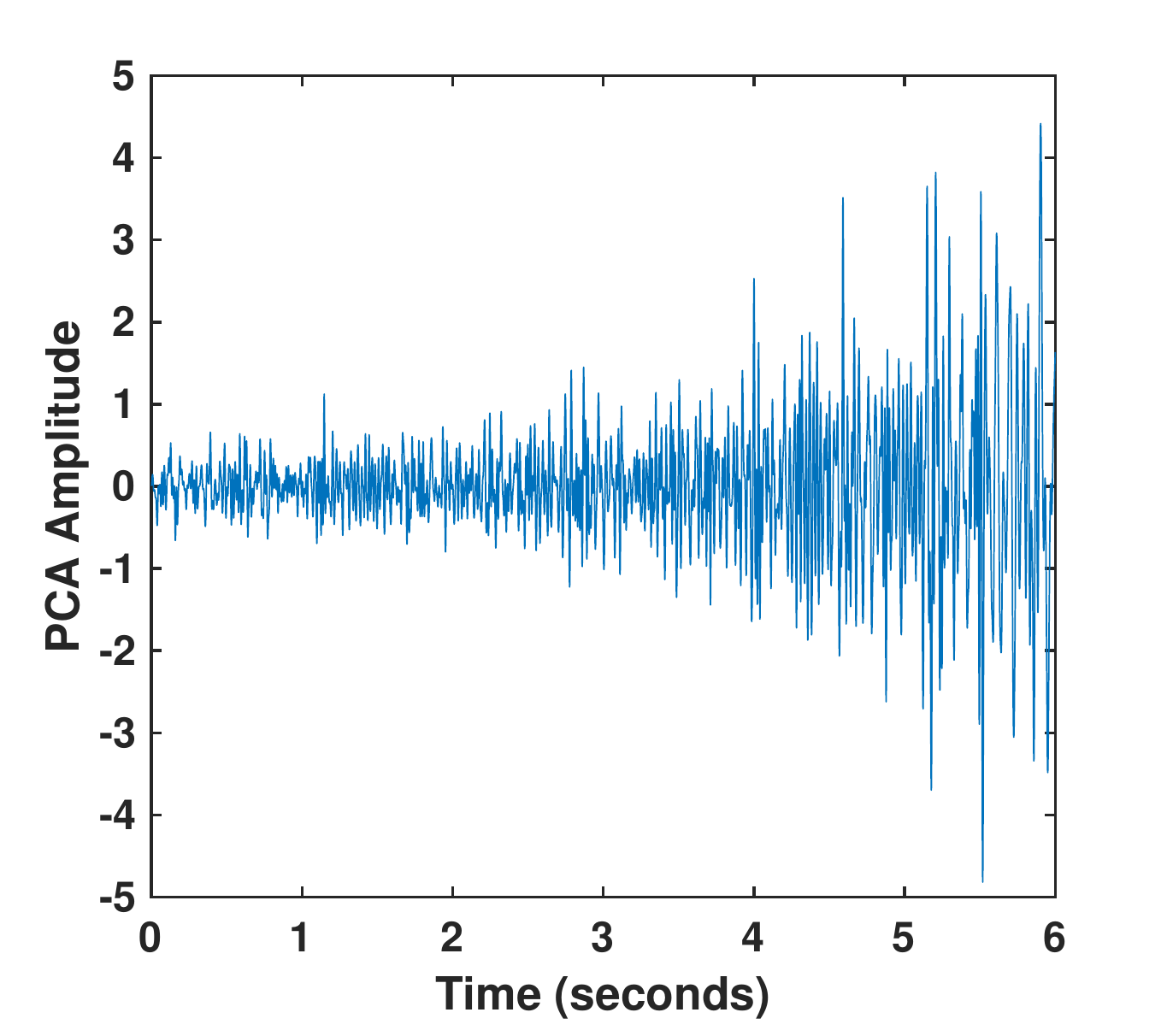}
			\label{fig_pca_d1_waveform}
		}
		\caption{Comparisons between the original CSI amplitudes of the 7\# subcarrier and the first PCA components of total 30 subcarriers for the two walking instances: (i) walking away from and (ii) walking towards the devices. (a) and (b) are separately the original CSI amplitude and the $1^{st}$ PCA component of instance (i); (c) and (d) are separately the original CSI amplitude and the $1^{st}$ PCA component of instance (ii).}
		\label{fig_walking_pca_waveform}
	\end{figure}

	As mentioned above, the time-series CSI amplitudes of 30 subcarriers within one CSI stream come from one Tx-Rx antenna pair, which means they reflect quite similar multipath propagation of Wi-Fi signals, and the amplitudes have correlated changing pattern.
	However, different CSI subcarriers have slightly different carrier frequencies, which results in some phase shifts and a little different attenuations of CSI amplitudes in each CSI stream.
	Directly using all the correlated data may push our system towards some deep fading and unreliable subcarriers, to ensure better recognition results, the system utilizes Principal Component Analysis (PCA) to automatically discover the correlation between CSI amplitudes in each CSI stream and produce synthesized combinations (principal components) \cite{wang2015understanding}.
	Fig. \ref{fig_walking_pca_waveform} displays the comparisons between original CSI amplitudes of the 7\# subcarrier and the first PCA components of total 30 subcarriers for the two walking instances mentioned above.
	We can see that the first PCA component can better depict the changing pattern and trend of the CSI variation induce by walking.
	Here the first three principal components, which capture the most details of walking movement, are selected as the refined CSI data.
	For each receiver, we can totally obtain 6 groups of refined CSI data sequences since there are 6 CSI streams.
	
	\subsection{Walking Profile Generation}
	\subsubsection{Walking Detection}
	Compared with other daily activities, such as sitting down and cooking, walking has some special characteristics: (i) it involves the motions of many different body parts, (ii) the moving speeds of different body parts are relatively high, (iii) it can last for a bit long time.
	Since the frequencies induced by walking are mainly between 20 Hz to 60 Hz, the Energy of Interest (EI), which equals the sum of (normalized) magnitudes squared of FFT coefficients in the frequency range 20$\sim$60 Hz, is calculated to detect walking movement in terms of an appropriate threshold like that in \cite{xu2018wistep,zeng2016wiwho}.
	The width of the FFT window is set as 256 ($\approx F_s/4$) based on the trade-off between detection accuracy and system response rate.
	Whenever a walking movement is detected, the system starts its core functions immediately, namely generating the walking profile and further recognizing walking gait and direction.
	
	\subsubsection{Spectrogram Generation}
\begin{figure}
		\centering
		\includegraphics[width=0.85\columnwidth]{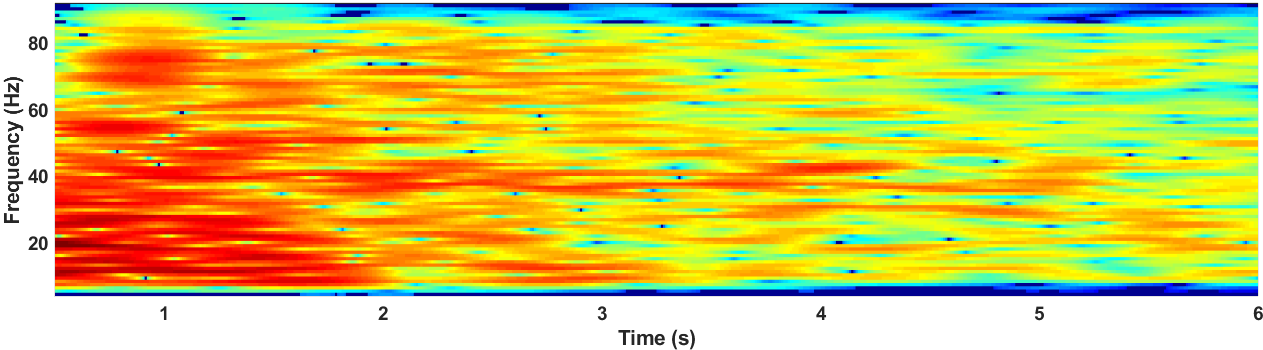}
		\caption{The spectrograms of the refined CSI data of a ``walking away" instance.}
		\label{fig_walking_away_pca_spectrogram}
	\end{figure}

	The refined CSI data can only depict the signal changing patterns in the time domain, where the signal reflections of different body parts are mixed together.
	When a subject is walking, its body parts (such as legs, arms and torso) have different moving speeds, and the signals reflected by different body parts have quite different energy considering different parts have different reflection area.
	Specifically, the swing leg (especially, the lower leg) has the highest moving speed, and the supporting leg and the torso have low moving speed; the signals reflected from the torso have the strongest energy while the arm reflections have much weak energy.
	By utilizing the Short-Time Fourier Transform (STFT), the system converts the time-domain refined CSI data sequence to the spectrogram of time-frequency domain.
	In practice, the CSI data sequences are first segmented into fixed-length chunks, which usually overlap each other so as to reduce artifacts at the boundary. 
	Then each chunk is transformed by FFT, and the logarithmic magnitudes squared of the FFT yield the final spectrogram.
	The spectrogram (of the $1^{st}$ PCA component) of the ``walking away" instance is illustrated in Fig. \ref{fig_walking_away_pca_spectrogram}, where the relatively ``hot" colored areas (have strong energy) in each chunk mainly present the torso reflections and some orange or yellow areas indicate the reflections of legs or arms.
	The time-varying trend of energy is still maintained in the spectrogram, which is a good evidence to judge the walking direction.
	Some advanced signal processing techniques used in \cite{wang2016gait}, like spectrogram enhancement, are not applied in our system to further refine the spectrogram, in contrary, we delegate more power to the RNN encoder-decoder network and let it learn to find out the important information from the noisy data.
	
	\subsubsection{Profile Splicing}
	Considering that original CSI amplitudes are denoised using a bandpass filter, the frequencies within [5, 90] (Hz) are reserved, and the frequency components induced by walking movement are mainly in the range of 20$\sim$60 Hz. 
	The proposed method keeps the data of 64 out of the 85 FFT points in the spectrogram, which corresponds to the frequency range of around 10$\sim$70 Hz, and stacks the data of each PCA component to form the primary \textit{walking profile} with dimensions $192 \times T$ for each PCA component, where $T$ is the length of the spectrogram (\textit{i.e.}, the amount of chunks).
	Thus the stacked 
	For enhancing the primary profile, the proposed system generates the dynamical features (also known as delta features) by applying discrete time derivatives on the primary profile, the dynamical features are proven to have excellent performance in many Automatic Speech Recognition (ASR) systems\cite{furui1986speaker,zheng2001comparison}, since the dynamical features can reveal some underlying connections and characteristics of adjacent speech frames.
	By concatenating the first-order derivative, namely the first-order difference, of the primary profile, the proposed method gets a $384 \times T$ dimensional walking profile for each CSI stream of a certain receiver.
	
	In addition, the proposed system is designed to splice the walking profiles corresponding to the same Tx-Rx antenna pair of the lowly-placed and the highly-placed receivers to further add more walking dynamics into the profile.
	Finally, a ``rich" and integrated walking profile with dimensions $768 \times T$ is constructed, and totally 6 integrated walking profile can be produced in each processing cycle.
	
	\subsubsection{Profile Reversion and Standardization}
	From Fig. \ref{fig_walking_pca_waveform}, we can find that the CSI amplitudes have relatively symmetrical time-varying patterns when the subject walks in reverse directions, which inspires us to reverse the data sequences of walking profiles along the time dimension to augment the instance data of the opposite walking movements.
	Usually, the performance of neural networks, like RNN, can be improved with more training data \cite{mikolov2012statistical}, in the proposed system the operation of profile reversion doubles the data used for training or recognition, and it is expected to promote the performance of our system.
	
	Before feeding data to the neural networks, it is necessary to standardize the data in advance, \textit{i.e.}, putting all the variables on the relatively same scale (with zero mean and unit variance), which can help to speed up the convergence and improve the performance of the networks \cite{shanker1996effect}.
	The statistical standardization method is applied in the system, to be specific, the proposed system calculates the global mean $\mu_{wp}$ and standard deviation $\sigma_{wp}$ of a integrated walking profile, and then subtracts $\mu_{wp}$ from each variable of the profile and subsequently divides the difference by $\sigma_{wp}$.
	
	So far, all the preparation work has been done, and it's time to build our attention-based RNN encoder-decoder networks.
	
	\subsection{Model Customization for Gait and Direction Recognition}
	The existing attention-based RNN encoder-decoder neural networks are specialized for Natural Language Processing (NLP) applications, especially for the machine translation task \cite{bahdanau2014neural}, and the trained networks can adaptively concentrate on important words in the source sentence when generating the target word.
	This distinguishing characteristic motivate us to create a cycle-independent gait and direction recognition system given the arbitrarily segmented CSI data.
	However, there are two main differences between the tasks of machine translation and ours.
	Firstly, machine translation is a single task learning problem, while our system aims to deal with two different tasks (direction and gait recognitions), namely multitask learning \cite{caruana1997multitask}.
	Secondly, apart from the source sentence, there exist statistical relations among the target words, which are basically described by statistical language model.
	Therefore, at the decoder side, the predicted target word is subsequently set as the input to predict the next word as illustrated in Fig. \ref{fig_attention_based_RNN_encoder_decoder}.
	However, in our system there is no explicit relation between human gait and walking direction, in order to jointly train the networks for our objectives, we need to customize our own model.
	\begin{itemize}
		\item \textit{\textbf{Encoder}}: In the system, the input of the RNN encoder is the CSI data sequence (the integrated $768 \times T$ dimensional walking profile), which is denoted as $\mathbf{x} = \{x^1, \cdots, x^T\}$, and each $x^t\ (t \in [1, T])$ is a vector with dimensions $768 \times 1$.
		The output of the RNN decoder is denoted as $\mathbf{y} = \{y^1, y^2\}$, where $y^1 \in \mathbb{R}^{1 \times n_d}$ and $y^2 \in \mathbb{R}^{1 \times n_g}$, $n_d$ and $n_g$ separately are the number of walking directions and the number of subjects.
		Moreover, we can define more variables for many other tasks such as gender classification.
		In this work, we mainly concern gait and direction recognition.	
		Considering that human gait and walking direction have no explicit relation with each other, the conditional probability of equation (\ref{eqn_attention_output_conditional_distribution}) needs to be rewritten as
		\begin{equation}\label{eqn_model_output_conditional_distribution}
		P(y^n| c^n) = g(h_y^n, c^n), n \in [1, 2] \mbox{,}
		\end{equation}
		and the computation of $h_y^n$ is also isolated to the former predicted target, namely $h_y^n = f(h_y^{n-1}, c^n)$.
		As suggested in \cite{bahdanau2014neural}, a bidirectional RNN (BiRNN) framework is utilized to create our RNN encoder.
		BiRNN presents each input sequence forwards and backwards to two separate recurrent hidden layers, which are connected to the same output layer, and it's reported to perform better than the unidirectional RNN \cite{schuster1997bidirectional,graves2012supervised}.
		In BiRNN, the hidden state $h_x^t$ is expressed as the concatenation of the forward hidden state $\overrightarrow{h_x^t}$ and the backward one $\overleftarrow{h_x^t}$, \textit{i.e.}, $h_x^t = [\overrightarrow{h_x^t}; \overleftarrow{h_x^t}]$.
		
		\item \textit{\textbf{Decoder}}: As for the decoder, because the computation of attention weight $\omega^{n,t}$ and attention vector $c^n$ doesn't depend on the predicted targets, which implies that the proposed method can directly and faithfully implement the computation processes of attention weights and attention vectors as described in \cite{bahdanau2014neural} without any major modification.
		The proposed method is expected to learn and compute valuable attention weights which enable the system to automatically align with some critical clips of the input data sequence for the two different tasks.
		Due to the lack of future context, a unidirectional RNN which has the same number of hidden layers as the encoder is employed in the proposed system.
		Thus, at the end of the encoding stage, the bidirectional hidden state $h_x^T$ of the last input of the encoder needs to be transformed to meet the size of the unidirectional hidden state $h_y^0$ of the decoder, and the transformation in this work is simply performing an additive operation between $\overrightarrow{h_x^T}$ and $\overleftarrow{h_x^T}$. 
		For the networks with multiple hidden layers, the transformation can be executed on each layer of the networks accordingly.
		Followed by the hidden layer(s), two full connection layers are adopted for the two different tasks.
	\end{itemize}
	
	\section{Implementation \& Evaluation}
	
	\subsection{Experiment Setup}
\begin{figure}[!t]
		\centering
		\includegraphics[width=0.8\columnwidth]{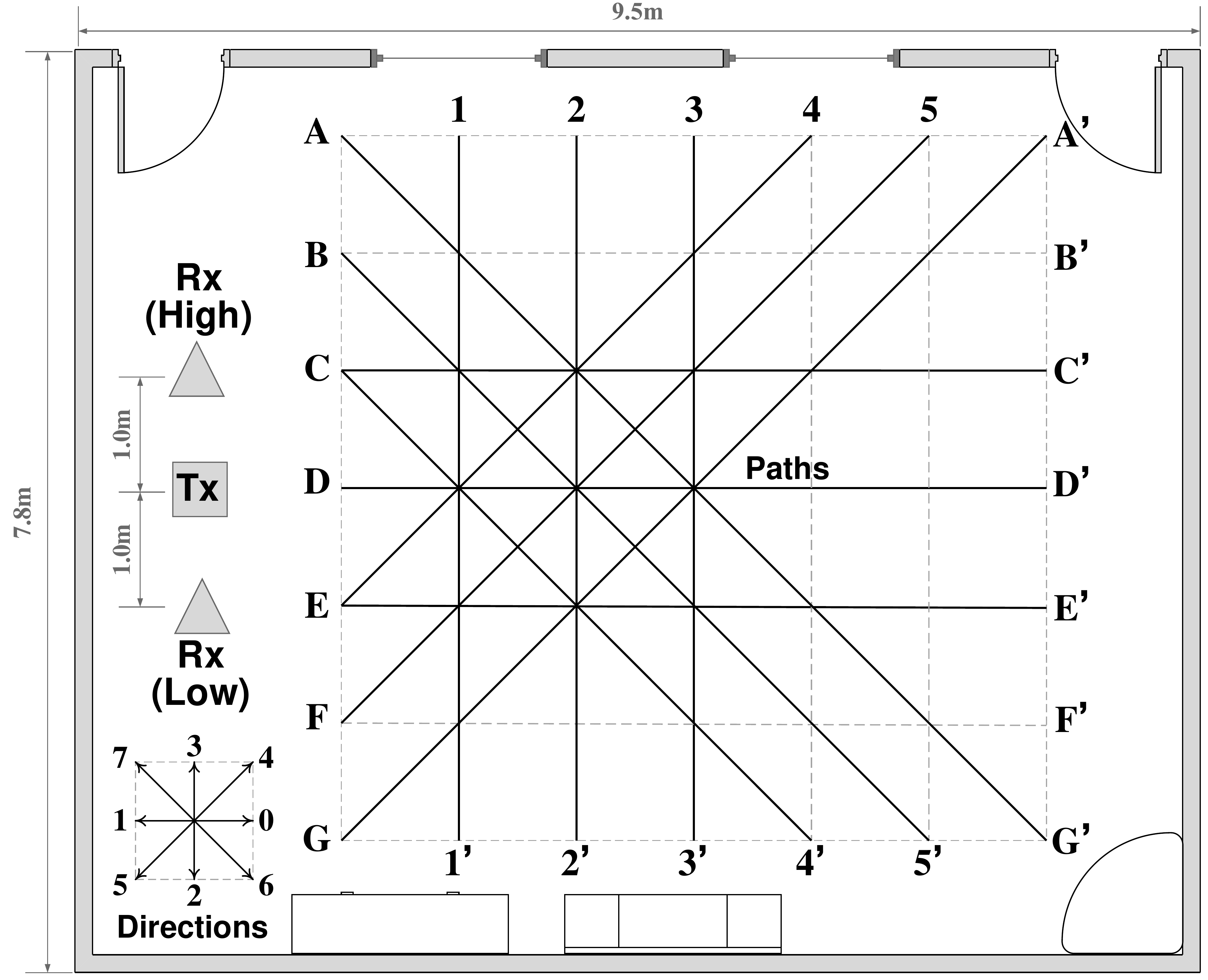}
		\caption{Experiment environment.}
		\label{fig_experiment_environment}
	\end{figure}

	In order to conveniently collect more CSI data of walking in different directions, we find a spacious laboratory with size 9.5m$\times$7.8m as the CSI data collection environment, which is shown in Fig. \ref{fig_experiment_environment}.
	The transmitter and the receivers are placed abreast at the marked positions and the distance between each receiver and the transmitter is 1.0 m.
	As we mentioned in subsection \ref{subsec_csi_collection}, for concentrating on different body parts of a walker, the two receivers are separately placed at heights of 0.5 m and 1.0 m above the ground level, and the transmitter is placed at a height of 0.75 m.
	The devices all run on the 149\# Wi-Fi channel (its central frequency is 5.745 GHz) without the dynamic frequency selection (DFS) and transmit power control (TPC) constraints of Federal Communications Commission (FFC).
	By using the CSI tool \cite{halperin2011tool}, the transmitter and the receivers are configured to work in monitor mode, where the data packets injected by the transmitter can simultaneously be captured by the two receivers.
	Besides, the clocks of the two receivers are synchronized by the Network Time Protocol (NTP) so as to obtain CSI measurements with synchronized timestamps.
	A deep learning server, which has two high-performance NVIDIA GeForce GTX 1080Ti Graphics Processing Units (GPUs), is connected to the receivers by cables.
	The server is specialized for data processing and model training.
	
	A 6.0m$\times$6.0m grid-layout area, 1.5 m away from the transmitter, is planned for the walking experiment, and the size of each grid is 1.0m$\times$1.0m.
	As shown in Fig. \ref{fig_experiment_environment}, 12 specific paths (marked with solid lines) are assigned for subjects to walk on, and the labels of 8 walking directions are annotated on the bottom left of the layout plan.
	The detailed experiment process is described in the next subsection.
	
	\begin{figure}[!t]
		\centering
		\includegraphics[width=0.65\columnwidth]{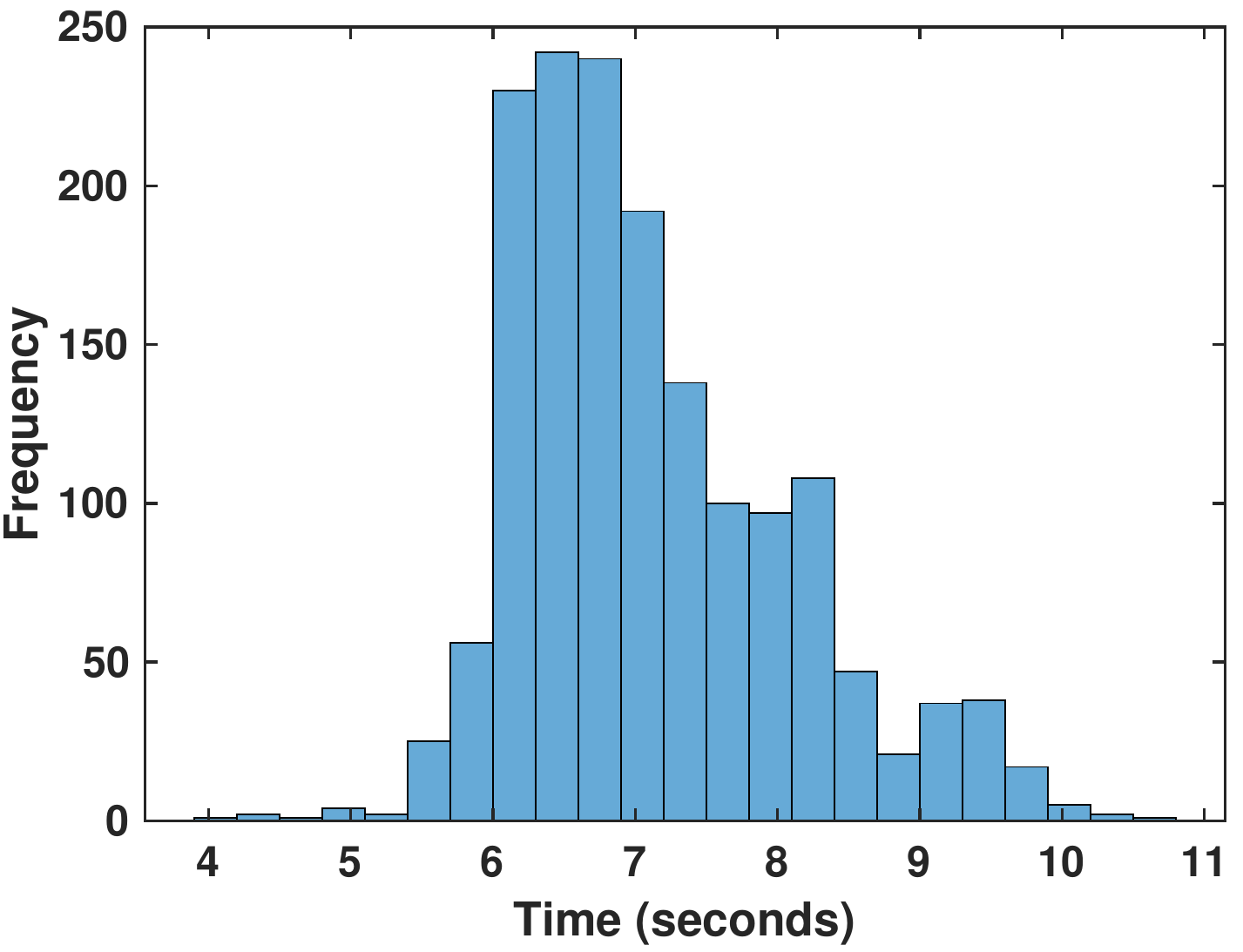}
		\caption{Time distribution of subjects walking on different paths.}	
		\label{fig_time_distribution}
	\end{figure}
	
	\subsection{Experiment Process}
	\subsubsection{Dataset Creation}
	In this work, 11 subjects (8 male and 3 female graduate students) are invited to carry on the walking experiment.
	For privacy concern, we don't record some private information like age, height and weight of each subject.
	Every time, a subject is asked to walk between the two endpoints of one straight path in its natural way, for example walking between D and D' of path DD' in Fig. \ref{fig_experiment_environment}.
	Meanwhile, each receiver sends its CSI measurements companied by timestamps to the server, and the server stores the CSI data and the timestamps for further processing.
	When the subject arrives at a endpoint, it turns around and walks back to the other endpoint.
	For each path, the subject is required to walk for 5 minutes without break (a alarm clock is provided to remind the subject), then turns to another path.
	Therefore, we can totally get 60-minute data involving 12 walking paths and 8 different directions from each subject.
	During the experiment, two video cameras are applied to record the whole experimental process like that in \cite{brajdic2013walk}, and the CSI data corresponding to walking (except for turning and break) are manually labeled by our recorders.
	We promise that all the video resources are only used for the experiment and will be cleared for protecting the privacy of subjects as soon as we finish the paper.
	
	An CSI data sequence, which involves a single-trip walking of a path, is regarded as an CSI walking instance, and we totally collect 10626 walking instances in the experiment.
	Since the shortest and the longest path lengths of the 12 paths are about 5.66 m and 8.49 m respectively, the time distribution of all subjects walking on different paths is illustrated in Fig. \ref{fig_time_distribution}, where the minimum and the maximum single-trip walking time separately is 4.135 s and 10.626 s.
	For each subject, we randomly select 20\% of the walking instances corresponding to a specific path and direction for validation and testing, and the remaining instances are for training.
	Given the 1000 Hz sampling rate of the system, for brevity, a 4000-point sliding window segmentation with step size of 500 points is performed on each labeled CSI walking instance to get multiple slices used for training, validation and testing.
	Note that thanks to the recurrent structure of RNN, the data with arbitrary lengths can be handled by our model.
	By conducting profile splicing and reversion, for training set, we get 172088 integrated walking profiles of all the 11 subjects.
	Moreover, we bisect the slices used for validation and testing and generate our validation and testing sets, each of which has 14858 profiles of all subjects.
	We randomly select the data of 8 subjects to train and evaluate our model.
	
	\subsubsection{Model Training}
	The training of our attention-based RNN encoder-decoder model is performed in PyTorch using one GTX 1080Ti GPU.
	Given the 11 GB memory size and the 11.3 TFLOPs (in single precision) processing power of an GTX 1080Ti GPU, it enables us to train a complex and relatively deep RNN encoder-decoder model.
	As a reference, a PyTorch implementation of the attten-based RNN encoder-decoder for machine translation \cite{bahdanau2014neural} is available on GitHub\footnote{https://github.com/spro/practical-pytorch}.
	
	In our specific implementation, the RNN architectures of our encoder and decoder both are GRU, which is reported to have simpler structure but better performance than LSTM \cite{jozefowicz2015empirical}, and the numbers of hidden layers are set as 3 in the encoder and decoder.
	The encoder and decoder of the proposed system have 1024 hidden units each, and the encoder has 256 input units while the decoder has no input layer.
	We use a minibatch stochastic gradient descent (SGD) algorithm to train the encoder and decoder, each SGD update direction is computed using a minibatch of 64 instances.
	We adjust (decrease) the learning rate from 1e-4 to 1e-6 based on the training epochs have been done, and the total training epochs of the system are set as 32.
	The validation set is employed to check if the error is within a reasonable range.
	For better preventing our model from overfitting, some effective techniques, such as dropout \cite{pham2014dropout} and training with noise \cite{an1996effects}, are introduced in the system.

	\subsubsection{Evaluation Metrics}
	For evaluating the performance of the proposed system, two major metrics are employed, namely $Accuracy$ and $F_1$ score, where $Accuracy$ is for primary evaluation of the system's gait and direction recognition ability since it only takes true positives of each class into consideration, and $F_1$ score, which is the weighted average of precision and recall, is used to evaluate the comprehensive performance of the system.
	Besides, the confusion matrices are also posted to illustrate the detailed recognition results.
	
	\begin{figure}[!t]
		\centering
		\subfigure[Human gait recognition.]
		{
			\includegraphics[width=0.45\columnwidth]{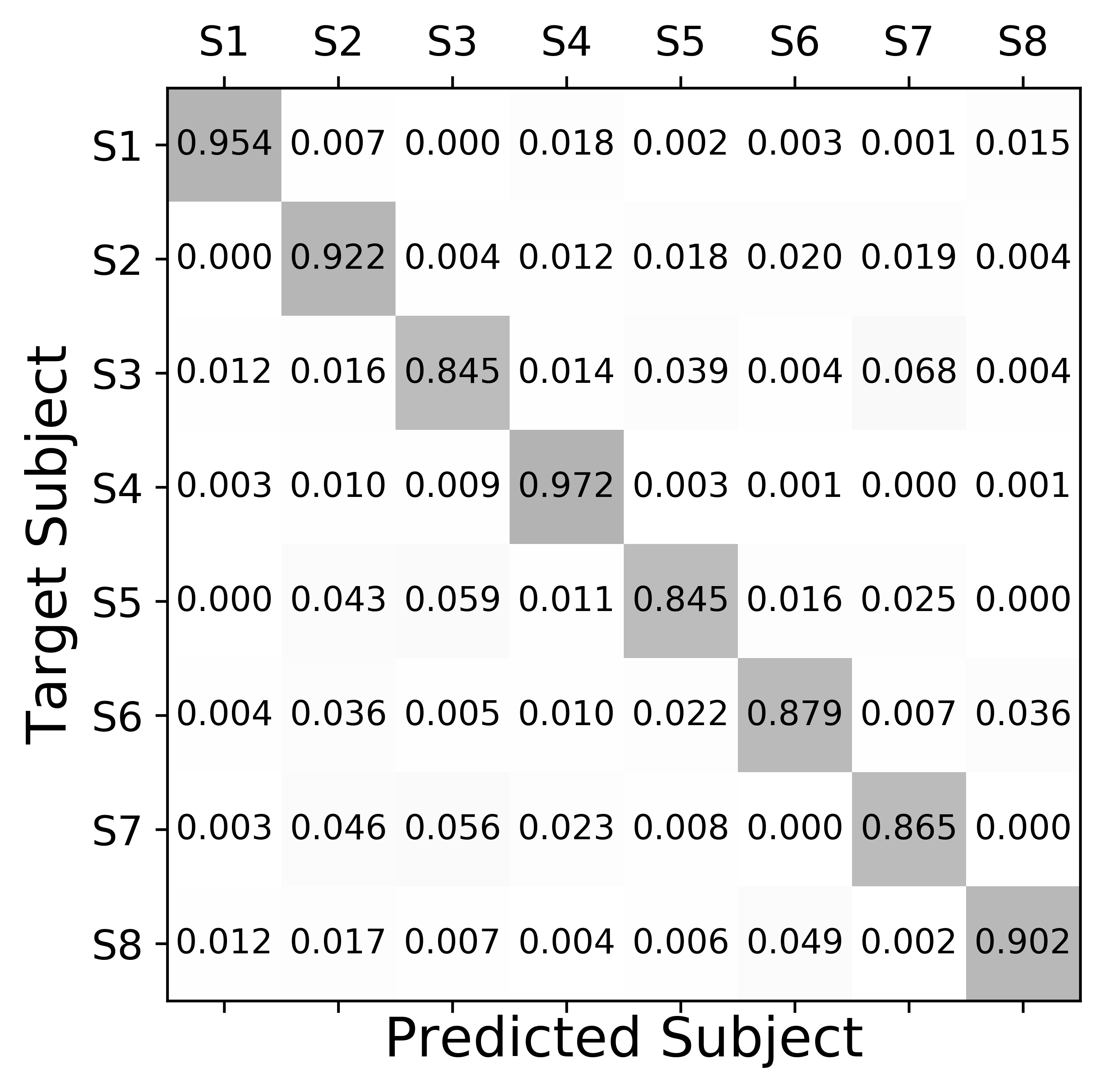}
			\label{fig_model_dir_cm}
		}\quad
		\subfigure[Walking direction recognition.]
		{
			\includegraphics[width=0.45\columnwidth]{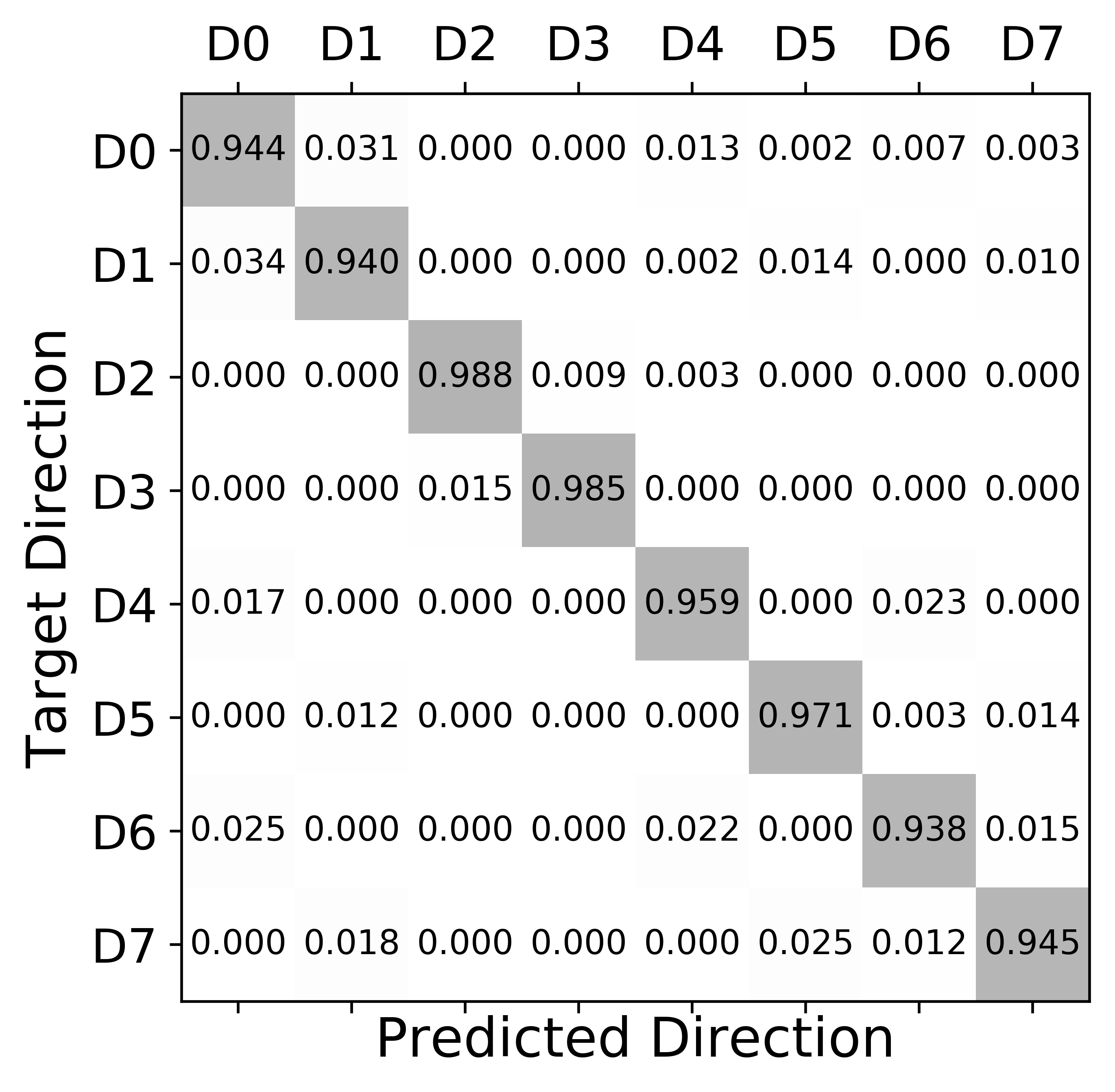}
			\label{fig_model_gait_cm}
		}
		\caption{Confusion matrices of gait and direction recognition results.}
		\label{fig_model_confusion_matrces}
	\end{figure}
	
	\begin{figure}[!t]
		\centering
		\subfigure[Human gait recognition.]
		{
			\includegraphics[width=0.65\columnwidth]{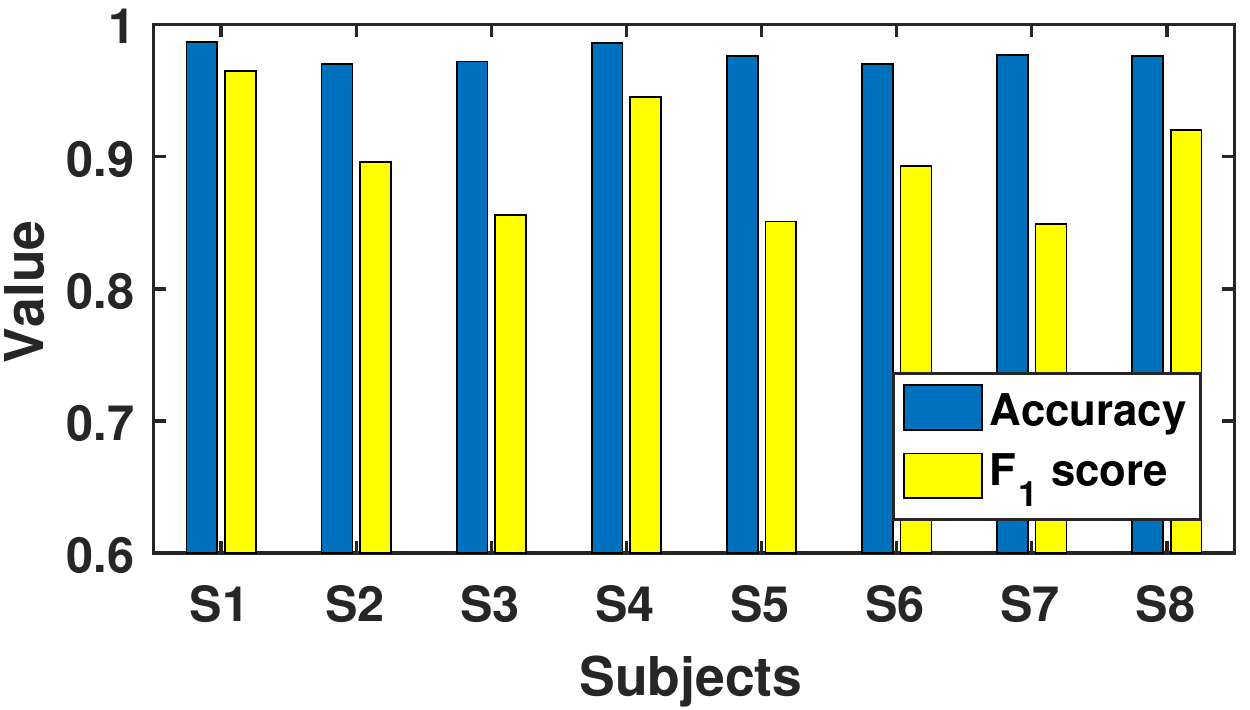}
			\label{fig_model_gait_result}
		}\quad
		\subfigure[Walking direction recognition.]
		{
			\includegraphics[width=0.65\columnwidth]{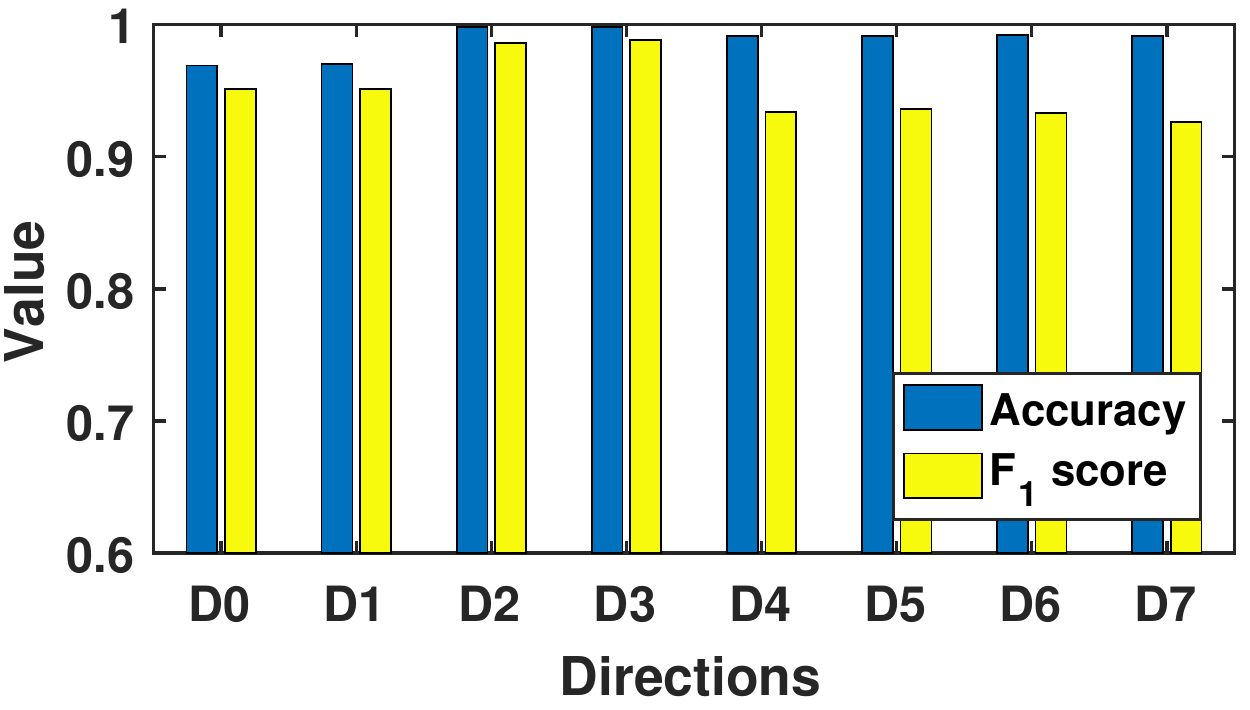}
			\label{fig_model_dir_result}
		}
		\caption{The detailed recognition accuracies and $F_1$ scores of human gait and walking direction.}
		\label{fig_model_dir_gait_results}
	\end{figure}
	
	\subsection{Experiment Results}
	The performance of the proposed system is evaluated on the testing dataset, and each instance in the test set has two labels, \textit{i.e.}, subject label (from ``S1'' to ``S8'') and direction label (from ``D0'' to ``D7'').
	The confusion matrices of human gait and walking direction recognition results are shown in Fig. \ref{fig_model_confusion_matrces}, where the diagonal entries represent the numbers of true positives of different classes, intuitively, we can find most of the instances are correctly predicted by the sytem.
	Fig. \ref{fig_model_dir_gait_results}, which is derived from the confusion matrices, illustrates the gait and direction recognition accuracies and $F_1$ scores for specific classes.
	The proposed system achieves relatively high accuracies (all above 95\%) and high $F_1$ scores of all the classes given different tasks, especially for the direction recognition task.
	Concretely, the average gait recognition accuracy and $F_1$ score of the 8 selected subjects are 97.68\% and 89.69\% respectively, while the average accuracy and $F_1$ score for direction recognition are separately 98.75\% and 95.06\%.
	These results demonstrate that by adopting the attention-based RNN encoder-decoder architecture can achieve promising recognition results on the human gait and walking direction recognition tasks.
	
	\begin{figure}[!t]
		\centering
		\subfigure[Instance of ``S8' and ``D1''.]
		{
			\includegraphics[width=0.45\columnwidth]{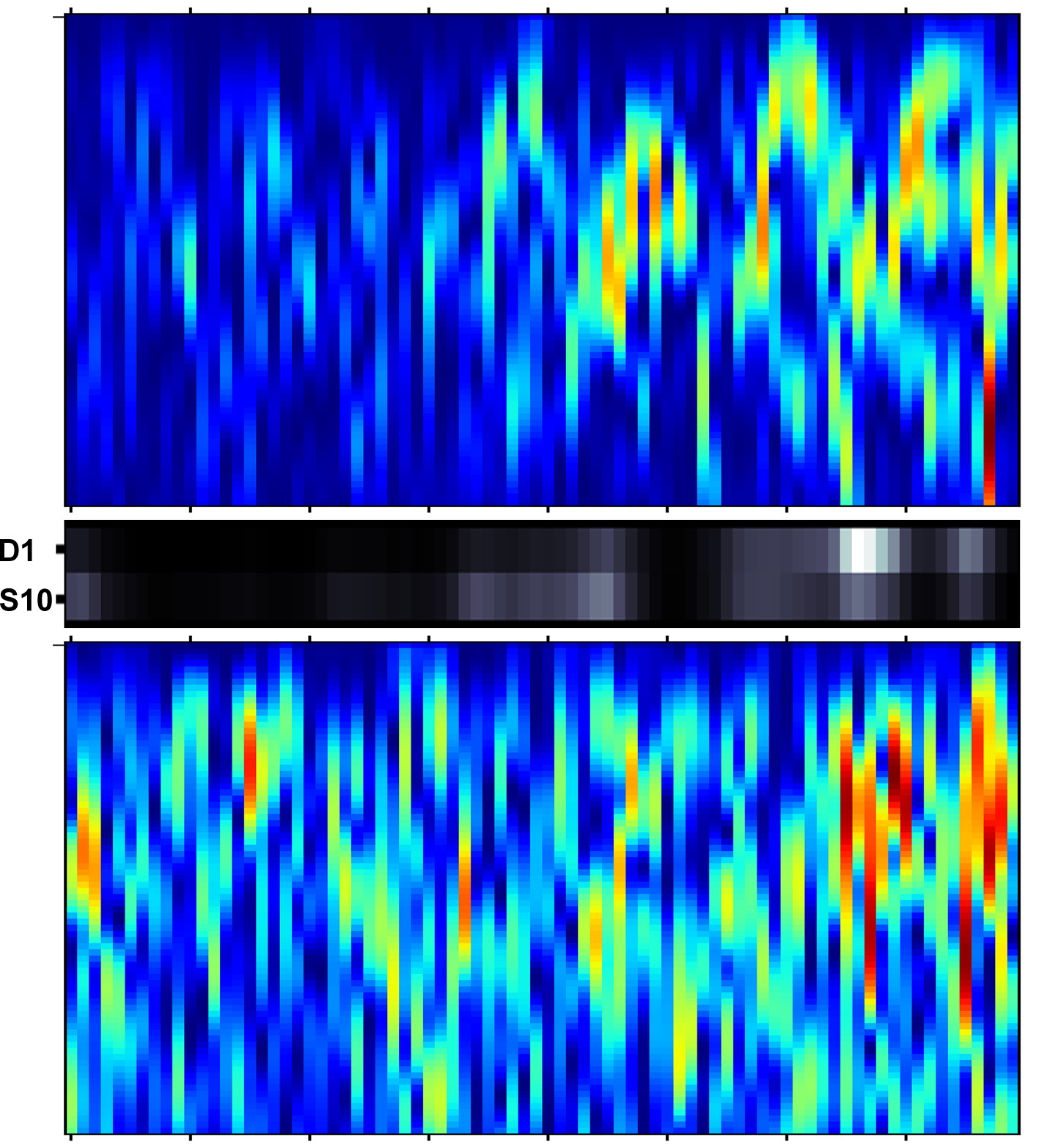}
			\label{fig_attention_weight_D1_S10}
		}
		\subfigure[Instance of ``S4'' and ``D4''.]
		{
			\includegraphics[width=0.45\columnwidth]{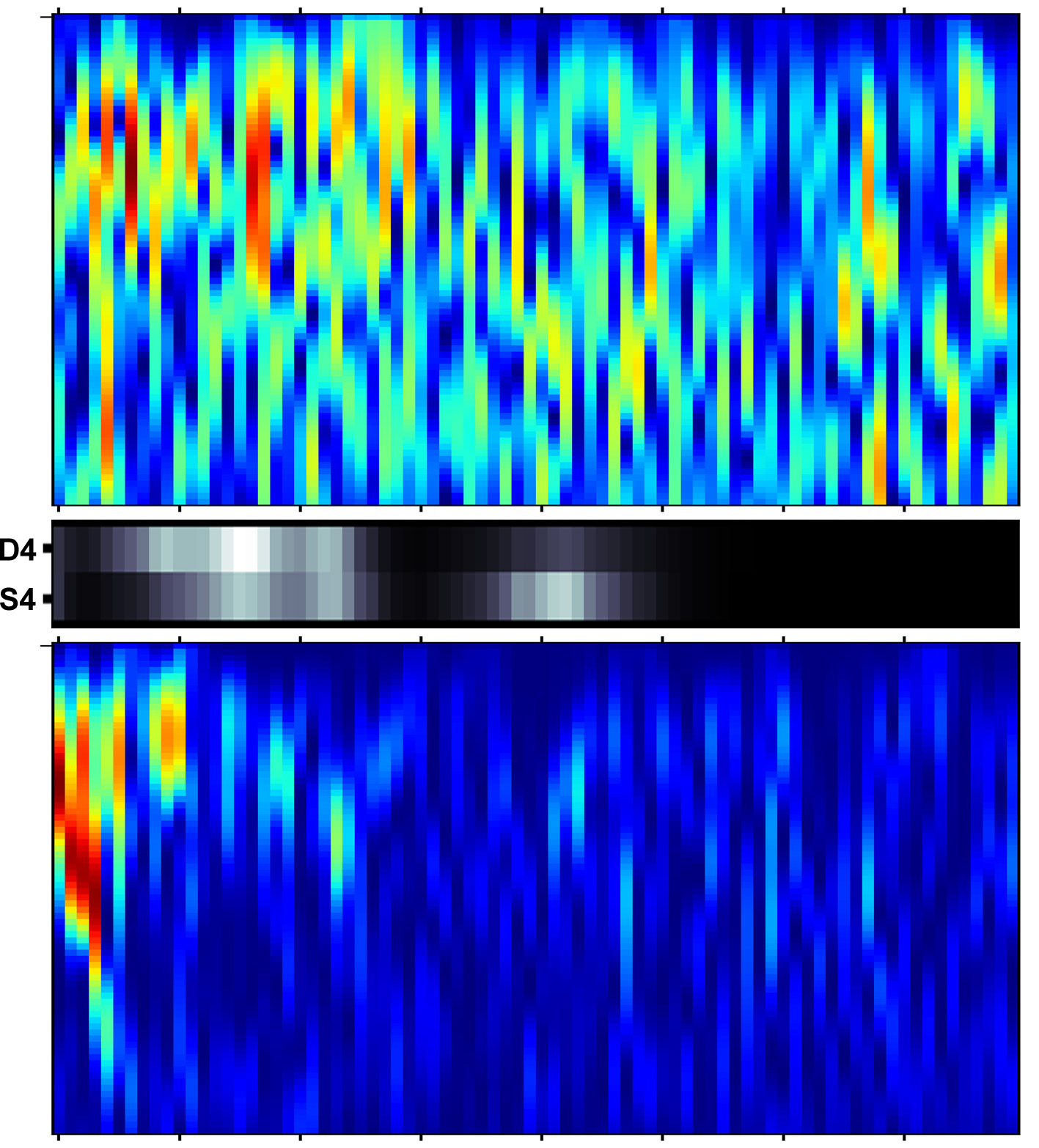}
			\label{fig_attention_weight_D4_S4}
		}
		\caption{Visualization of attention weights for 2 test instances.}
		\label{fig_attention_weight}
	\end{figure}
	
	\subsection{Attention Visualization}
	Given a certain test instance, the proposed system first encodes the instance into a particular attention vector, which is the weighted sum of all the hidden states of the system's encoder, then the system decodes the attention vector and outputs the prediction for a specific task.
	The attention weights computed by the equation \ref{eqn_attention_weight} score how well the walking profile at each time step and the current prediction match.
	Fig. \ref{fig_attention_weight} visualizes the attention weights (in the middle of each subfigure, where the upper row and the lower row correspond to direction and gait respectively) of 3 test instances in grayscale (where 0 is black and 1 is white), and the top and bottom parts of each subfigure are the spectrograms of the highly-placed Rx (HighRx) and lowly-placed Rx (LowRx), respectively.
	We observe that (i) for different recognition tasks, the computed weights are different, which means the proposed system can automatically focus its attentions on different time steps of the spectrograms when coping with different tasks; (ii) the large weights (more brighter weights) are basically align with some high-energy clips in the spectrograms, where these clips usually contain much more critical and apparent features of human walking dynamics.
	Based on the observations, we can conclude that even without portioning CSI data sequence into cycle-wise time slices the proposed system can still learn to adaptively align with some important parts of the sequence and realize cycle-independent gait and walking direction recognition.

	\section{Limitations}
	Although we have proven the feasibility and shown the promising results of jointly recognizing human walking gait and direction with the attention-based RNN encoder-decoder framework in Wi-Fi networks.
	There are still some limitations.
	We only evaluate the performance of our system on a group of 8 subjects, and some external influence factors, like footwear, floor surface, room layout, and internal influence, like length of walking profile, haven't taken into consideration.
	We are trying to explore the impacts of subject group size and other specific factors in our future work.
	Limited by the bandwidth and synchronization problem of Wi-Fi networks, if there are multiple people walking at the same time, the signals reflected off different people are mixed together at the receiver side.
	As many existing WiFi-based systems, for now, we haven't found some effective methods to separate the mixed Wi-Fi signals from multiple moving individuals, and now the proposed system can't be applied in the scenario where multiple individuals walk simultaneously.
	
	\section{Conclusions}
	By adopting the attention-based RNN encoder-decoder framework, we proposed a new cycle-independent human gait and walking direction recognition system which was jointly trained for walking gait and direction recognition purposes.
	For capturing more human walking dynamics, two receivers and one transmitter were deployed in different spatial layouts.
	The CSI measurements from the two receivers were first gathered together and refined to form an integrated walking profile.
	Then, the RNN encoder read and encoded the walking profile into primary feature vectors, based on which the decoder computed different attention vectors for different recognition tasks.
	With the attention scheme, the proposed system could learn to adaptively align with different critical clips of the CSI data sequence for walking gait and direction recognitions.
	We implemented our system on commodity Wi-Fi devices in indoor environment, and the experimental results demonstrated that the proposed system could achieve average $F_1$ scores of  89.69\% for gait recognition from a group of 8 subjects and 95.06\% for direction recognition from 8 directions, besides the average accuracies of these two recognition tasks both exceeded 97\%.
	Our system is expected to enable more practical and interesting applications.


\end{document}